\PassOptionsToPackage{prologue,dvipsnames}{xcolor}
\documentclass{article}

\usepackage{microtype}
\usepackage{graphicx}
\usepackage{subfigure}
\usepackage{booktabs}
\usepackage{hyperref}
\usepackage{enumitem}

\usepackage[accepted]{icml2023}

\usepackage{amsmath}
\usepackage{amssymb}
\usepackage{mathtools}
\usepackage{amsthm}
\usepackage{bm}
\usepackage{color,soul} %
\usepackage{mathrsfs}
\usepackage[capitalize,noabbrev]{cleveref}

\theoremstyle{plain}
\newtheorem{theorem}{Theorem}[section]

\theoremstyle{definition}

\theoremstyle{remark}

\usepackage{subfigure}
\usepackage{amsfonts}
\usepackage{nicematrix}
\usepackage{array}
\usepackage{multirow}
\usepackage{tikz}
\usepackage{enumitem}
\usepackage{subcaption}
\usepackage{siunitx}
\usepackage{adjustbox}
\newcolumntype{P}[1]{>{\raggedright\arraybackslash}p{#1\textwidth-2\tabcolsep-1.5\arrayrulewidth}}
\newcolumntype{Y}[1]{>{\centering\arraybackslash}p{#1\textwidth-2\tabcolsep-1.5\arrayrulewidth}}
\usepackage[textsize=tiny]{todonotes}

\icmltitlerunning{Inverse Design in Distributed Circuits Using Single-Step Reinforcement Learning}

\usepackage{amsmath,amsfonts,bm}

\def\eqref#1{equation~\ref{#1}}

\def\1{\bm{1}}

\def\va{{\bm{a}}}

\def\vp{{\bm{p}}}

\def\vs{{\bm{s}}}

\def\vu{{\bm{u}}}

\def\mI{{\bm{I}}}

\def\mR{{\bm{R}}}

\DeclareMathAlphabet{\mathsfit}{\encodingdefault}{\sfdefault}{m}{sl}
\SetMathAlphabet{\mathsfit}{bold}{\encodingdefault}{\sfdefault}{bx}{n}

\def\gA{{\mathcal{A}}}
\def\gB{{\mathcal{B}}}

\def\gF{{\mathcal{F}}}

\def\gH{{\mathcal{H}}}

\def\gL{{\mathcal{L}}}
\def\gM{{\mathcal{M}}}

\def\gP{{\mathcal{P}}}

\def\gS{{\mathcal{S}}}

\def\gY{{\mathcal{Y}}}

\newcommand{\E}{\mathbb{E}}

\begin{document}

\twocolumn[
\icmltitle{Inverse Design in Distributed Circuits Using Single-Step Reinforcement Learning}

\icmlsetsymbol{equal}{*}

\begin{icmlauthorlist}
\icmlauthor{Jiayu Li}{equal,yyy}
\icmlauthor{Masood Mortazavi}{comp}
\icmlauthor{Ning Yan}{comp}
\icmlauthor{Yihong Ma}{equal,sch}
\icmlauthor{Reza Zafarani}{yyy}
\end{icmlauthorlist}

\icmlaffiliation{yyy}{Syracuse University, New York, USA}
\icmlaffiliation{comp}{Futurewei Technologies Inc., California, USA}
\icmlaffiliation{sch}{University of Notre Dame, Indiana, USA}

\icmlcorrespondingauthor{Jiayu Li}{jli221@data.syr.edu}
\icmlcorrespondingauthor{Masood Mortazavi}{masood.mortazavi@futurewei.com}

\icmlkeywords{Machine Learning, ICML}

\vskip 0.3in
]

\printAffiliationsAndNotice{\icmlEqualContribution} %

\begin{abstract}
The goal of inverse design in distributed circuits is to generate near-optimal designs that meet a desirable transfer function specification. 
Existing design exploration methods use some combination of strategies involving artificial grids, differentiable evaluation procedures, and specific template topologies.
However, real-world design practices often require non-differentiable evaluation procedures, varying topologies, and near-continuous placement spaces.
In this paper, we propose DCIDA, a design exploration framework that learns a near-optimal design sampling policy for a target transfer function. DCIDA decides all design factors in a compound single-step action by sampling from a set of jointly-trained conditional distributions generated by the policy. 
Utilizing an injective interdependent ``map", DCIDA transforms raw sampled design ``actions" into uniquely equivalent physical representations, enabling the framework to learn the conditional dependencies among joint ``raw'' design decisions.
Our experiments demonstrate DCIDA's Transformer-based policy network achieves significant reductions in design error compared to state-of-the-art approaches, with significantly better fit in cases involving more complex transfer functions.
\end{abstract}

\section{Introduction}
\begin{figure}
	\centering
	\subfigure[A 4-resonator circuit]{
		\begin{minipage}{0.37\linewidth}
			\includegraphics[width=1\linewidth]{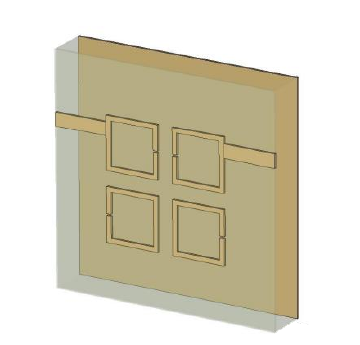} 
		\end{minipage}
		\label{f1a}
	}
     \hspace{1cm} %
    	\subfigure[Physical representation]{
    		\begin{minipage}{0.38\linewidth}
   		 	\includegraphics[width=1\linewidth]{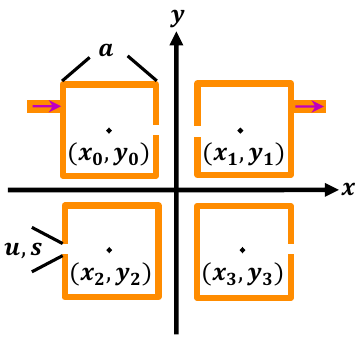}
    		\end{minipage}
		\label{f1b}
    	}
	\caption{4-resonator circuit and its physical representation.}
	\label{f1}
\end{figure}

As 5G and 6G communication technologies and quantum computing advance towards higher frequencies, it becomes important for distributed circuit designs on resonators to achieve desirable performance~\cite{he2020deep, deory2024low}. These designs often include components such as square resonators~\cite{073c4e9db8e14bae9ed4a1eeff116a79} shown in Figure \ref{f1}, ring resonators~\cite{1a65ae7432db4ef6b4e4d013069a5383}, or bar resonators~\cite{989962}. The specification of these circuits is expressed as a transfer function, which describes the frequency behavior of the distributed circuit. 

Designing a distributed circuit to achieve a desired transfer function remains a slow and laborious process. To accelerate the process, Circuit-GNN~\cite{pmlr-v97-zhang19e} trains a graph neural network (GNN) as a pre-trained model to predict the transfer function $s_{21}(\{\omega_{i}\}^{m}_{i=1})$ of a given design, which represents the ratio of output-to-input signals for frequencies $\{\omega_{i}\}^{m}_{i=1}$. By using the differentiability of this pre-trained model, Circuit-GNN solves the inverse design problem for $s_{21}$, but it relies on specific candidate topologies. In fact, experts often lack a predetermined preference for topology types. Additionally, predictions of $s_{21}$ from the pre-trained model are less accurate than those generated by electromagnetic (EM) simulators. This gives rise to a key challenge: \emph{How can we generate distributed circuits that meet a desired $s_{21}$ without assuming targeted topologies while enabling the use of simulation?} 

For our design space exploration (DSE) problem, we can adopt reinforcement learning (RL)---a DSE approach deployed in various domains~\cite{gate_sizing, rl_eda_case_studies, con_clocks, PhysRevFluids.6.053902}. 
Prior research~\cite{Jiang2021DelvingIM, mirhoseini2020chip, 10.1109/2022.3185540,FENGERDSE} has applied RL to DSE problems utilizing traditional multi-step RL to sample and identify near-optimal design decisions. 
The multi-step approach encodes partial design states and employs non-analytic gridding and masking techniques to satisfy conditionality on action probabilities. 
However, partial designs cannot provide performance feedback, the artificially imposed grids reduce fidelity, and output masking eliminate feedback opportunity. Single-step $\theta$-Resonance~\cite{mortazavi2022thetaresonance} over-came these limitations for definite-horizon fixed-dimensional design explorations. It avoided artificial gridding pitfalls that often grip multi-step RL DSE but relied on heuristic penalty assignment to the unavoidable unconditioned anomalous designs, and it only explored categorical design spaces.

To address the above limitations, we propose \underline{D}istributed \underline{C}ircuit \underline{I}nverse \underline{D}esign \underline{A}utomation (DCIDA), an inverse design framework tailored to generate near-optimal distributed circuits. 
DCIDA employs a trainable policy network that transforms a constant input (representing the blank design slate) to a set of conditional distributions defined on design dimensions. 
The corresponding joint distribution can be used to sample a series of discrete and continuous actions as a compound action in a single step. 
We introduce implied boundaries and deterministic interdependent mapping functions that map raw design actions to the physical representations of the distributed resonators.
Our experiments show DCIDA outperforms Circuit-GNN and $\theta$-Resonance by generating distributed circuits that achieve significant reduction in error. 
DCIDA obtains the results without targeted topologies or the number of resonators. 
In summary, our main contributions include:
\begin{itemize}
    \item We propose DCIDA---an inverse design framework trained by single-step RL to generate near-optimal distributed circuits that meet desired transfer function $s_{21}$ more tightly than Circuit-GNN.
    \item We introduce design boundaries to reduce the DSE complexity. Within the boundaries, we formulate deterministic mapping functions that accurately translate raw compound actions into physical circuit designs.
    \item Our experiments show that DCIDA excels in generating distributed circuits that more tightly meet desired $s_{21}$, significantly outperforming Circuit-GNN and $\theta$-Resonance.
\end{itemize}

\section{Related Work}
\noindent{\textbf{Automating lumped circuit design.}} Lumped circuits consist of lumped components such as resistors, capacitors and transistors. Research in this domain focuses on determining optimal device parameters and circuit topologies to meet desired specifications. Some approaches employ optimization-based methods, including Bayesian Optimization \cite{pmlr-v80-lyu18a}, Geometric Programming \cite{1249422}, and Genetic Algorithms \cite{5699917, 10.1145/2330163.2330318}, while some studies propose learning-based methods including supervised learning methods~\cite{dong2023cktgnn} and RL methods \cite{10.5555/3408352.3408464, 10.5555/3437539.3437740, 10.1145/3489517.3530501}. \emph{However, the methods for lumped circuit design are not applicable to high-frequency circuits with distributed design.}

\noindent{\textbf{Automating high-frequency circuit design.}}
In the distributed circuit designs~\cite{5290124, 7847393}, one state-of-the-art method is Circuit-GNN \cite{pmlr-v97-zhang19e}, which first trains a graph neural network as a model to predict the transfer function $s_{21}$. Next, in inverse design exploration, Circuit-GNN produces circuits to meet desired transfer functions, using corresponding templates specifying topology and the number of resonators (corresponding to the target transfer functions in the test set). 
Recent work RLDFCDO~\cite{gao2024automated} picks one circuit from all candidate circuits and tunes the selected circuit using RL with proximal policy optimization (PPO) algorithm~\cite{schulman2017proximal}. \emph{The methods are different from ours. They mainly aim to predict the EM properties of distributed circuits or rely on candidate template (i.e., topology types) to solve the inverse design problem.}

\section{Preliminary}
\label{pre}
\begin{figure*}
    \centering
    \includegraphics[width=\linewidth]{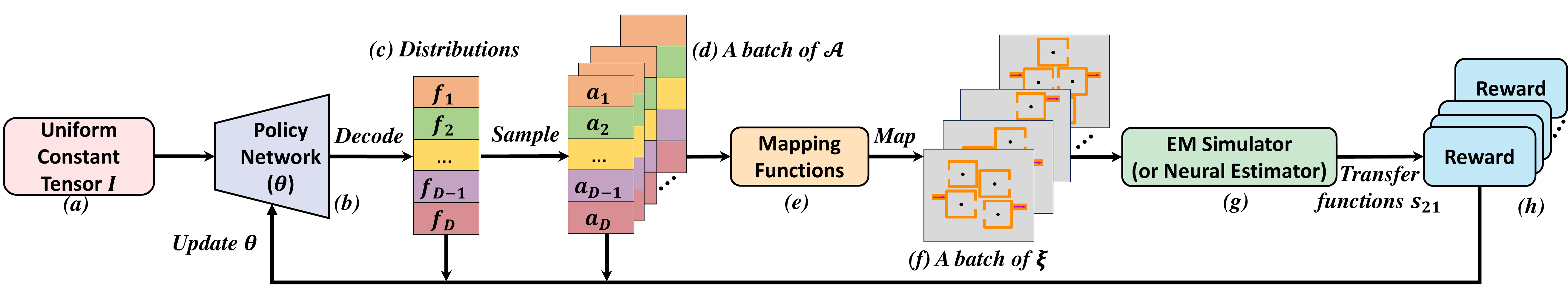}
        \caption{Overview of DCIDA. A policy network (b) decodes constant input tensor $\mI$ (a) into conditional distributions (c). From these distributions, a batch of compound action $\gA$ (d) are sampled. These ``raw'' samples are mapped to their uniquely corresponding physical representations of circuits (f) through injective mapping functions (e). We examine the circuits (f) with an EM simulator, or neural estimator, (g) in order to obtain the corresponding transfer function for each design representation. These transfer functions' $L_1$ distances from target transfer function yield the rewards (h) used to update $\theta$ according to \ref{policy_optimization}.}
    \label{f2}
\end{figure*}
The properties of a square resonator include length $a \in \mathbb{R}$, its center $p = \{(x,y)| x\in \mathbb{R}, y\in \mathbb{R}\}$ on a plane, and the direction of an open slit. The slit can face one of four directions: up, down, left, or right. We can use a one-hot vector $u \in \{0,1\}^{4}$ to represents the direction of a slit, where a single entry is $1$ and others are $0$. The position of a slit relates to a vector $s \in \mathbb{R}^{4}$, where the non-zero entry corresponds to the slit direction, and the remaining entries are zero. A distributed circuit with $N$ resonators is defined by the tuple $(\vp, \va, \vu, \vs)$, where $\vp \in \mathbb{R}^{N\times2}$, $\va \in \mathbb{R}^{N}$, $\vu \in \{0,1\}^{N\times4}$ and $\vs \in \mathbb{R}^{N\times4}$. An EM simulator evaluates the distributed circuit by its transfer function $\hat{s}_{21} = \hat{\gY}(\vp, \va, \vu, \vs)$ where  $\hat{\gY} \in \mathbb{C}^{m}$ and $\mathbb{C}$ is a complex number. We compute the error $\epsilon_\text{db}$ between $\hat{\gY}$ and the target transfer function $\gY \in \mathbb{C}^{m}$ in db domain
\begin{equation}
\epsilon_\text{db}=\frac{20}{m}\sum^{m}_{i=1}|\log(|\gY_{i}|)-\log(|\hat{\gY}_{i}|)|.
\end{equation}
For clarity in the paper, we focus on the distributed circuit design using square resonators as the blocks. However, our approach is general and can be used with other resonators.

\section{Inverse Design of Distributed Circuit}
\subsection{Problem Formulation}
As illustrated in Figure \ref{f2}, the DCIDA framework $\gH$ has trainable parameters $\theta$ and uses a constant variable $\mI$ as an input. Given the number of resonators $N$ and a desired transfer function $s_{21}$ without candidate templates, DCIDA learns to decode $\mI$ to generate a near-optimal distributed circuit design of form $(\vp, \va, \vu, \vs)$.
The objective is to minimize the error $\epsilon_\text{db}$ between the generated circuit's transfer function $\hat{s}_{21}$ and the desirable transfer function $s_{21}$. 

As noted earlier, we treat the generation of distribute circuits as a DSE problem. Since no intermediate error for partial designs can be computed, we formulate our DSE problem (with a definite fixed number of decisions) as a Single-Step Markov Decision Process (SSMDP), $s_0 \rightarrow \gA \rightarrow s_D$, in terms of compound action $\gA = \{a_1, a_2, ... , a_D\}$. Here, $s_0 \in \gS$ and $s_D \in \gS$ represent the initial blank slate and the complete design (i.e., $s_D=(\vp, \va, \vu, \vs)$), respectively. 

Parameters $\theta$ determining the sampling policy network $\pi$ are tuned to minimize a statistical loss defined by our RL algorithm described in section \ref{policy_optimization}.
\begin{equation}
\small
\label{target}
\theta^{+} \sim \underset{\theta}{\text{min}} \quad \gL(\theta) \quad \rightarrow \quad \pi_{\theta^{+}} \sim \pi_{\theta^{*}}\in \Pi^{*}(\gS, \gA, R),
\end{equation}
Here, the SSMDP is given by $(\gS, \gA, R)$, $\Pi^{*}$ is the set of potential optimal policies in response to the SSMDP dynamics with the optimal $\theta^{*}$, and $\gL$ corresponds to the objective function related to penalty $\epsilon_\text{db}$ and defined by our RL algorithm. Parameters $\theta^{+}$ defines a near optimal policy.

\subsection{Problem Solution}
\subsubsection{Learning Joint Distributions and Inter-Dependencies}
To solve the SSMDP problem, we introduce a RL framework using a policy network with parameters $\theta$ to generate a sampling policy $\pi_{\theta} = \pi_{\theta}(a_1,...,a_D|\mI)$ where $\mI \in \mathbb{R}^{d}$ is the initial state $s_0$ with ones in tensor form. The $s_D$ is terminal state achieved by a compound action $\gA$ sampled from $\pi_\theta$. 

\noindent \textbf{Policy Network.} Given $N$ number of resonators, the policy network with $\mI$ produces a policy demonstrated by a jointly mixed actions distribution, which can be represented in the conditional form of probability density functions $\gF$ in Eq.(\ref{eq1}).\vspace{-2mm}
\begin{equation}
\small
\begin{split}
\label{eq1}
    \pi_{\theta}(\gA|\mI) &=\prod_{i=1}^{D}f_{i}(a_{i}|a_{i+1}\cdot\cdot\cdot a_{D};\mI, \theta), 
\end{split}
\end{equation} 
where we sample the $i$-th design action $a_i$ from $f_{i} \in \gF$. $\gF$ include beta distributions for sampling bounded continuous actions and categorical distributions for sampling discrete actions. We experiment with MLP and Transformer as policy networks in section \ref{experiment}. The number of resonators $N$ decides the number of actions $D$ shown in section \ref{mapping}.

\begin{figure*}
    \centering
    \includegraphics[width=0.7\linewidth]{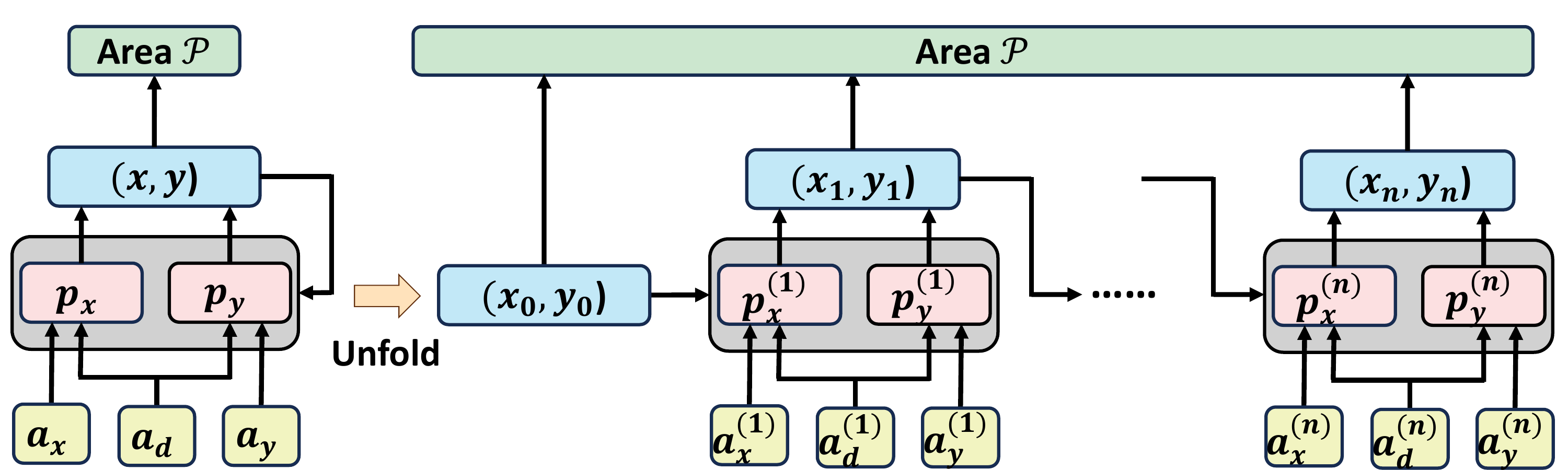}
    \caption{The cascade of injective interdependent functions determines the centers of resonators within the area $\gP \subseteq \gB$.} 
    \label{f3}
\end{figure*}
\subsubsection{Mapping Actions into Designs}
\label{mapping}
We design boundaries and formulate a series functions to map actions into physical representations $(\vp, \va, \vu, \vs)$ within the boundaries in the circuit space $\gM$. 

\noindent \textbf{Slit position $(u, s)$ and resonator length $a$.} A discrete action $a_u\in \{0,1,2,3\}$ specifies a slit’s direction, where $0$, $1$, $2$, and $3$ correspond to up, left, down, and right, respectively. We use the action $a_u\in \{0,1,2,3\}$ to decide a slit's direction $u \in \{0,1\}^{4}$ as follows,
\begin{equation}
\small	
u(i)=\left\{
\begin{array}{rcl}
1, \quad i=a_u\\
0, \quad i\neq a_u
\end{array} \right.
\end{equation}
After determining a slit's direction, a continuous action $a_s\in[0,1]$ decides the slit's position $s$ relative to the center of the corresponding edge: 
\begin{equation}
\small	
s(i) = 
\begin{cases} 
\frac{1}{8} \tanh(2a_s - 1), & i = a_u \\ 
0, & i \neq a_u 
\end{cases}
\end{equation}
Since all resonators have the same length, a resonator's length, ranging from $L$ to $2L$, is given by $a = L\times(a_l+1)$, where an action $a_l \in [0,1]$ is the corresponding action.

\noindent \textbf{Resonator position $p$.} Starting the DSE on a blank slate  leads to an infinitely large exploration space, making precise placement of resonators inefficient. To address the issue, we propose the Theorem \ref{th1} to restrict exploration space within predefined boundaries. \textbf{\emph{Proofs of the Theorems are shown on the Appendix \ref{theorems_proof}.}}

\begin{theorem}
\label{th1}
In the circuit space $\gM$, given the resonator length $a$, the maximum gap $g = ag_\text{max}$ (with a predefined $g_\text{max}$), and the number of resonators $N$, we define an area $\gP = \{(x, y) \mid x \in [0, B - a], y \in [\frac{a - B}{2}, \frac{B - a}{2}]\}$, where $B = aN + g(N-1)$, as the region bounding the centers of square resonators in a distributed circuit. The distributed circuit is further confined to the area $\gB = \{(x, y) \mid x \in [-\frac{a}{2}, \frac{2B - a}{2}], y \in [-\frac{B}{2}, \frac{B}{2}]\}$, where $\gP \subseteq \gB$.
\end{theorem}

\noindent \textbf{The offsets about positions of resonators.} To precisely determine the position of each resonator within the boundary $\gB$, we set the default center for the leftmost (first) resonator as $(0, 0)$. Subsequently, starting from the first resonator, we sequentially add remaining resonators from left to right based on the position of the last placed resonator. Using the actions $a_{f} \in \{0,1,2\}$, $a_{us} \in [0, 1]$ and $a_{ug} \in [0, 1]$ to adjust the offsets, we compute the shift factor as follows:\vspace{-1mm}
\begin{equation}
\small	
f=
\begin{cases}
0, & a_f = 0\\
0.2, & a_f=1\\
0.5, & a_f=2
\end{cases}
\end{equation}
The uniform shift factor is $u_s = a_{us}$, and the uniform gap factor is $u_g = a_{ug}$. We define the deviation function for the shift as $d_s(l,r,u_{s}) = u_{s}(r-l) + l$, where $l = 0$ and $r=af$. The deviation function for the gap is $d_g(l,r,u_{g})=e^{u_{g}(\log{\frac{r}{l}}) + l}$, where $l=ag_\text{min}$ and $r=ag_\text{max}$. Here, $g_\text{min}$ and $g_\text{max}$ are the predefined minimum and maximum ratio of the gap.

\noindent \textbf{The interdependent functions.} We propose interdependent functions to accurately locate the position of each resonator by using the deviation functions. Starting with the default center of the first resonator, we define Definition \ref{th2} and Definition \ref{th3} to construct the functions. The functions map actions $a^{(i)}_{x} \in [0, 1]$ and $a^{(i)}_{y} \in [0, 1]$ for a resonator $i$ to the point $p = (x_{i}, y_{i})$ where $p \in \gP$. $(x_{i}, y_{i})$ is a center of the resonator $i$. An action $a^{(i)}_{d} \in \{0,1,2\}$ corresponds to a relative position (i.e., up, down and right respectively) of the current resonator relative to the previous one. Since mapping the rightmost resonator differs from that of the other resonators, we denote the rightmost resonator as resonator $n$ and remaining resonators (except the leftmost one) as resonator $i$, where $1\leq i<n$. Figure \ref{f3} shows the structure of the interdependent functions. Based on the mapping method of $N$ resonators, the total number of actions is $D=8N-5$. 
\begin{theorem}
\label{th2}
Given the center of the last resonator $(x_{i-1},y_{i-1})$, the interdependent functions $p^{(i)}_{x}: a^{(i)}_{x} \rightarrow x_{i}$ and $p^{(i)}_{y}: a^{(i)}_{y} \rightarrow y_{i}$ map actions $\{a^{(i)}_{x}, a^{(i)}_{y}, a^{(i)}_{d}\}  \in \gA$ to the point $p = (x_{i}, y_{i})$ as the center of the current resonator $i$.

\resizebox{.75\linewidth}{!}{
\begin{minipage}{\linewidth}
\begin{align}
\label{eq20}
h^{(i)}_{x}&=
\begin{cases}
a^{(i)}_{x}\min(x_{i-1}+d_s, B-2a-d_g)+(1-a^{(i)}_{x})x_{i-1}, \qquad \qquad \quad & \text{$a^{(i)}_{d} = 0$}\\
a^{(i)}_{x}\min(x_{i-1}+d_s, B-2a-d_g)+(1-a^{(i)}_{x})x_{i-1}, \qquad \qquad \quad &\text{$a^{(i)}_{d} = 1$}\\
\min(x_{i-1}+a+d_g, B-2a-d_g), \qquad \qquad \qquad \qquad \qquad \quad & \text{$a^{(i)}_{d} = 2$}\\
\end{cases}\\
h^{(i)}_{y}&=
\begin{cases}
\min(y_{i-1}+a+d_g, \frac{B-a}{2}), &\text{$a^{(i)}_{d} = 0$}\\
\max(y_{i-1}-a-d_g, \frac{a-B}{2}), &\text{$a^{(i)}_{d} = 1$}\\
a^{(i)}_{y}\min(y_{i-1}+d_s, \frac{B-a}{2}) + (1 - a^{(i)}_{y})\max(y_{i-1}-d_s, \frac{a-B}{2}), &\text{$a^{(i)}_{d} = 2$}\\
\end{cases}
\end{align}
\end{minipage}
}
\end{theorem}

\begin{theorem}
\label{th3}
Given the actions $\{a^{(n)}_{x}, a^{(n)}_{y}, a^{(n)}_{d}\} \in \gA$ for the rightmost resonator $n$ and the center of the previous resonator $(x_{n-1},y_{n-1})$, the interdependent functions $p^{(n)}_x: a^{(n)}_{x} \rightarrow x_{n}$ and $p^{(n)}_{y}: a^{(n)}_{y} \rightarrow y_{n}$ compute the center $p=(x_{n}, y_{n})$ of the rightmost resonator $n$.

\resizebox{.75\linewidth}{!}{
\begin{minipage}{\linewidth}
\begin{align}
\label{eq21}
h^{(n)}_x&=
\begin{cases}
a^{(n)}_{x}\min(x_{n-1}+d_s, B-a)+(1-a^{(n)}_{x})x_{n-1},  \quad \quad \quad \ \quad \ \quad \quad \quad &\text{$a^{(n)}_{d} = 0$}\\
a^{(n)}_{x}\min(x_{n-1}+d_s, B-a)+(1-a^{(n)}_{x})x_{n-1},   \quad \quad \quad \ \quad \ \quad \quad \quad &\text{$a^{(n)}_{d} = 1$}\\
\min(x_{n-1}+d_g+a, B-a), \qquad \qquad \qquad \qquad \qquad \quad& \text{$a^{(n)}_{d} = 2$}\\
\end{cases}\\
h^{(n)}_{y}&=
\begin{cases}
\min(y_{n-1}+a+d_g, \frac{B-a}{2}), &\text{$a^{(n)}_{d} = 0$}\\
\max(y_{n-1}-a-d_g, \frac{a-B}{2}) &\text{$a^{(n)}_{d} = 1$}\\
a^{(n)}_{y}\min(y_{n-1}+d_s, \frac{B-a}{2})+(1 - a^{(n)}_{y})\max(y_{n-1}-d_s, \frac{a-B}{2}), &\text{$a^{(n)}_{d} = 2$}\\
\end{cases}
\end{align}
\end{minipage}
}
\end{theorem}
Figure \ref{f3} shows the workflow of the interdependent functions. In Appendix \ref{case_study1}, we show an example of determining centers of resonators within the boundary $\gP$ using the sampled actions and the mapping functions. \vspace{-2mm}

\begin{table}[t]
\renewcommand\arraystretch{1.3}
\centering
  \caption{\underline{Error $\epsilon_\text{db}$} of inverse designs on circuits with 3 and 4 resonators. We report the average performance of our method, Circuit-GNN and $\theta$-Resonance. The best results are in \textbf{bold}. The reduction of error are from comparisons between our method and the best results of baselines.}
   \vspace{-0.1in}
   \label{tab:performance1}
  \resizebox{\linewidth}{!}{
  \LARGE
  \begin{NiceTabular}{>{\centering\arraybackslash}p{2.2cm}:>{\centering\arraybackslash}p{2.2cm}:>{\centering\arraybackslash}p{2.8cm}:>{\centering\arraybackslash}p{2.8cm}:>{\centering\arraybackslash}p{4.7cm}}
    \toprule
    \#-resonator circuits & Topology Type & $\theta$-Resonance (db) & Circuit-GNN (db) & Transformer-based DCIDA (db) (\textbf{ours}) \\
    \midrule
    \multirow{4}{*}{3} 
     & 0 & 5.73 \Large{$\pm 2.24$} & 1.12 \Large{$\pm$ 0.49} & \textbf{0.61} \Large{$\pm$ 0.27} ($\downarrow$ -45.54)\\
     & 1 & 4.12 \Large{$\pm 2.00$} & 0.96 \Large{$\pm$ 0.37} & \textbf{0.70} \Large{$\pm$ 0.37} ($\downarrow$ -27.08)\\
     & 2 & 5.60 \Large{$\pm 2.44$} & 0.84 \Large{$\pm$ 0.39} & \textbf{0.55} \Large{$\pm$ 0.15} ($\downarrow$ -34.52)\\
     & 3 & 4.60 \Large{$\pm 3.15$} & 1.40 \Large{$\pm$ 0.64} & \textbf{0.99} \Large{$\pm$ 0.53} ($\downarrow$ -31.25)\\
    \cline{1-5}
    \multirow{10}{*}{4}
     & 0 & 4.65 \Large{$\pm 2.47$} & 2.12 \Large{$\pm$ 1.02} & \textbf{0.69} \Large{$\pm$ 0.42} ($\downarrow$ -66.82)\\
     & 1 & 3.81 \Large{$\pm 2.19$} & 1.47 \Large{$\pm$ 0.53} & \textbf{0.85} \Large{$\pm$ 0.50} ($\downarrow$ -42.17)\\
     & 2 & 6.86 \Large{$\pm 1.92$} & 1.18 \Large{$\pm$ 0.42} & \textbf{1.14} \Large{$\pm$ 0.53} ($\downarrow$ -3.39)\\
     & 3 & 4.26 \Large{$\pm 2.26$} & 1.30 \Large{$\pm$ 0.43} & \textbf{0.91} \Large{$\pm$ 0.33} ($\downarrow$ -30.00)\\
     & 4 & 5.36 \Large{$\pm 2.80$} & 1.34 \Large{$\pm$ 0.49} & \textbf{1.18} \Large{$\pm$ 0.51} ($\downarrow$ -11.94)\\
     & 5 & 5.15 \Large{$\pm 2.50$} & 1.31 \Large{$\pm$ 0.19} & \textbf{0.88} \Large{$\pm$ 0.29} ($\downarrow$ -32.82)\\
     & 6 & 6.18 \Large{$\pm 2.96$} & 1.77 \Large{$\pm$ 0.63} & \textbf{1.19} \Large{$\pm$ 0.65} ($\downarrow$ -32.77)\\
     & 7 & 4.39 \Large{$\pm 2.68$} & 1.54 \Large{$\pm$ 0.56} & \textbf{0.95} \Large{$\pm$ 0.41} ($\downarrow$ -38.31)\\
     & 8 & 7.16 \Large{$\pm 2.51$} & 1.27 \Large{$\pm$ 0.26} & \textbf{0.96} \Large{$\pm$ 0.37} ($\downarrow$ -24.41)\\
     & 9 & 4.71 \Large{$\pm 2.39$} & 1.28 \Large{$\pm$ 0.42} & \textbf{0.93} \Large{$\pm$ 0.47} ($\downarrow$ -27.34)\\
    \bottomrule
  \end{NiceTabular}}
  \vspace{-0.15in}
\end{table}

\subsubsection{Policy Optimization} 
\label{policy_optimization}
With the variation on PPO algorithm~\cite{schulman2017proximal}, we optimize the design sampling policy $\pi_{\theta}$, which can generate samples of compound action $\gA$. Our proposed deterministic mapping strategy (see Section \ref{mapping}) maps each sample into its corresponding physical representation $\xi=(\vp, \va, \vu, \vs)$, which is then input to an EM simulator (or neural estimator) to produce transfer functions $\hat{\gY}(\xi)$. Based on the target transfer function $\gY$, we compute a sample reward $R(\xi)$.
\begin{equation}
\small
\label{eq3}
R(\xi)=
-\epsilon_\text{db}=-\frac{20}{m}\sum^{m}_{i=1}|\log(|\gY_{i}|)-\log(|\hat{\gY}_{i}(\xi)|)|.
\end{equation}
After obtaining the rewards for the current batch of $n$ samples, $\mR=\{R(\xi_{1}),...,R(\xi_{n})\}$, we use the renewal rate $\alpha_{r}$ to update the ``running'' reward ${\hat{R}}$ in the $t$-th iteration as follows:
\begin{equation}
\small
    \hat{R}_{t} = \alpha_{r}\E_{\xi\sim\pi_{\theta}}[\mR]+ (1-\alpha_{r})\hat{R}_{t-1}.
\end{equation}
In contrast to PPO, we use the running reward $\hat{R}_{t}$ to unbias the sample rewards and compute the sample advantage for compound action $\gA$.
\begin{equation}
\small	
    \hat{A}^{\pi_{\theta}}_{t}(\gA, \mI) = R-\hat{R}_{t}.
\end{equation} 
We define policy objective function of DCIDA based on PPO's unclipped surrogate objective function, incorporating the sum of the reversed Kullback-Leibler (KL) distance of $f_{i}$ as a surrogate KL-regularizer and entropy H of $f_{i}$ as surrogate entropy regularizer to promote exploration~\cite{schulman2017proximal, pmlr-v97-ahmed19a, dbLP:journals/corr/abs-2009-10897}. \vspace{-1mm}
\begin{equation}
\small
\begin{split}
\label{loss}
\gL(\theta) &= -\mathop{\E}_{\gA\sim\pi_{\theta_\text{old}}}[\frac{\pi_{\theta}(\gA|\mI)}{\pi_{\theta_\text{old}}(\gA|\mI)}\hat{A}^{\pi_{\theta_\text{old}}}_t(\gA, \mI)]\\&+\sum^{D}_{i=1}(\beta_\text{KL}\text{KL}(f_{i}(\theta)|| f_{i}(\theta_\text{old}))-\beta_{e,t}\text{H}(f_{i}(\theta))),
\end{split}
\end{equation} 
where $f_{i}(\theta)=f_{i}(a_{i}|a_{i+1}\cdot\cdot\cdot a_{D};\mI, \theta)$. We have $\beta_\text{KL}$ and $\beta_{e,t}=\max(\beta_\text{min}, \beta_{e,t-1} \cdot \beta_\text{decay})$ to control exploration.

\noindent \textbf{Training Algorithm.} In each iteration of training, we batch $n$ design samples, partition the batch into $z$ mini-batches and then set $E$ epochs for training a batch. With the objective $\gL(\theta)$, we apply stochastic gradient descent to perform the backpropagation and update parameters $\theta$ in the policy network through $\frac{nE}{z}$ loops, which updates $\pi_{\theta}$ towards an optimal policy $\pi_{\theta^{*}}$.
\begin{table}[t]
\renewcommand\arraystretch{1.3}
\centering
\caption{\underline{Error $\epsilon_\text{db}$} of inverse designs on circuits with 5 and 6 resonators. We use the same settings as Table \ref{tab:performance1}, and report the average performance of our method, Circuit-GNN and $\theta$-Resonance.}
  \vspace{-0.1in}
  \label{tab:performance2}
  \resizebox{\linewidth}{!}{
  \LARGE
   \begin{NiceTabular}{>{\centering\arraybackslash}p{2.2cm}:>{\centering\arraybackslash}p{2.2cm}:>{\centering\arraybackslash}p{2.8cm}:>{\centering\arraybackslash}p{2.8cm}:>{\centering\arraybackslash}p{4.7cm}}
    \toprule
    \#-resonator circuits & Topology Type & $\theta$-Resonance (db) & Circuit-GNN (db) & Transformer-based DCIDA (db) (\textbf{ours}) \\
    \midrule
    \multirow{9}{*}{5}
     & 0 & 4.12 \Large{$\pm 3.02$} & 1.08 \Large{$\pm$ 0.35} & \textbf{1.07} \Large{$\pm$ 0.29} ($\downarrow$ -0.93)\\
     & 1 & 3.56 \Large{$\pm 2.22$} & 1.31 \Large{$\pm$ 0.39} & \textbf{1.30} \Large{$\pm$ 0.41} ($\downarrow$ -0.76)\\
     & 2 & 3.69 \Large{$\pm 2.47$} & 1.12 \Large{$\pm$ 0.31} & \textbf{1.09} \Large{$\pm$ 0.26} ($\downarrow$ -2.68)\\
     & 3 & 5.39 \Large{$\pm 1.35$} & 1.15 \Large{$\pm$ 0.28} & \textbf{1.06} \Large{$\pm$ 0.38} ($\downarrow$ -7.83)\\
     & 4 & 2.86 \Large{$\pm 2.83$} & 1.27 \Large{$\pm$ 0.47} & \textbf{0.96} \Large{$\pm$ 0.49} ($\downarrow$ -24.41)\\
     & 5 & 5.09 \Large{$\pm 2.58$} & 1.29 \Large{$\pm$ 0.45} & \textbf{0.99} \Large{$\pm$ 0.22} ($\downarrow$ -23.26)\\
     & 6 & 5.41 \Large{$\pm 2.35$} & 1.17 \Large{$\pm$ 0.39} & \textbf{1.15} \Large{$\pm$ 0.35} ($\downarrow$ -1.71)\\
     & 7 & 3.82 \Large{$\pm 3.10$} & 1.43 \Large{$\pm$ 0.42} & \textbf{1.14} \Large{$\pm$ 0.42} ($\downarrow$ -20.28)\\
     & 8 & 3.41 \Large{$\pm 2.47$} & 1.38 \Large{$\pm$ 0.43} & \textbf{1.21} \Large{$\pm$ 0.43} ($\downarrow$ -12.32)\\
     \cline{1-5}
    \multirow{6}{*}{6}
     & 0 & 3.22 \Large{$\pm 2.51$} & 1.55 \Large{$\pm$ 0.56} & \textbf{1.23} \Large{$\pm$ 0.61} ($\downarrow$ -20.65)\\
     & 1 & 3.64 \Large{$\pm 1.85$} & 1.72 \Large{$\pm$ 0.62} & \textbf{1.29} \Large{$\pm$ 0.60} ($\downarrow$ -25.00)\\
     & 2 & 5.20 \Large{$\pm 1.97$} & 1.36 \Large{$\pm$ 0.39} & \textbf{1.17} \Large{$\pm$ 0.31} ($\downarrow$ -13.97)\\
     & 3 & 4.05 \Large{$\pm 2.40$} & 1.40 \Large{$\pm$ 0.42} & \textbf{1.18} \Large{$\pm$ 0.58} ($\downarrow$ -15.71)\\
     & 4 & 4.54 \Large{$\pm 3.02$} & 1.79 \Large{$\pm$ 0.35} & \textbf{1.16} \Large{$\pm$ 0.52} ($\downarrow$ -35.20)\\
     & 5 & 4.80 \Large{$\pm 2.27$} & 1.29 \Large{$\pm$ 0.38} & \textbf{1.15} \Large{$\pm$ 0.40} ($\downarrow$ -10.85)\\
    \bottomrule
  \end{NiceTabular}}
  \vspace{-0.15in}
\end{table}

\section{Experimental Evaluations}
\label{experiment}
\begin{figure*}[t]
\centering
\subfigure[3-resonator circuit and its transfer function from \textbf{DCIDA}]{
\begin{minipage}{0.5\linewidth}
\centering
\resizebox{\linewidth}{!}{\includegraphics[]{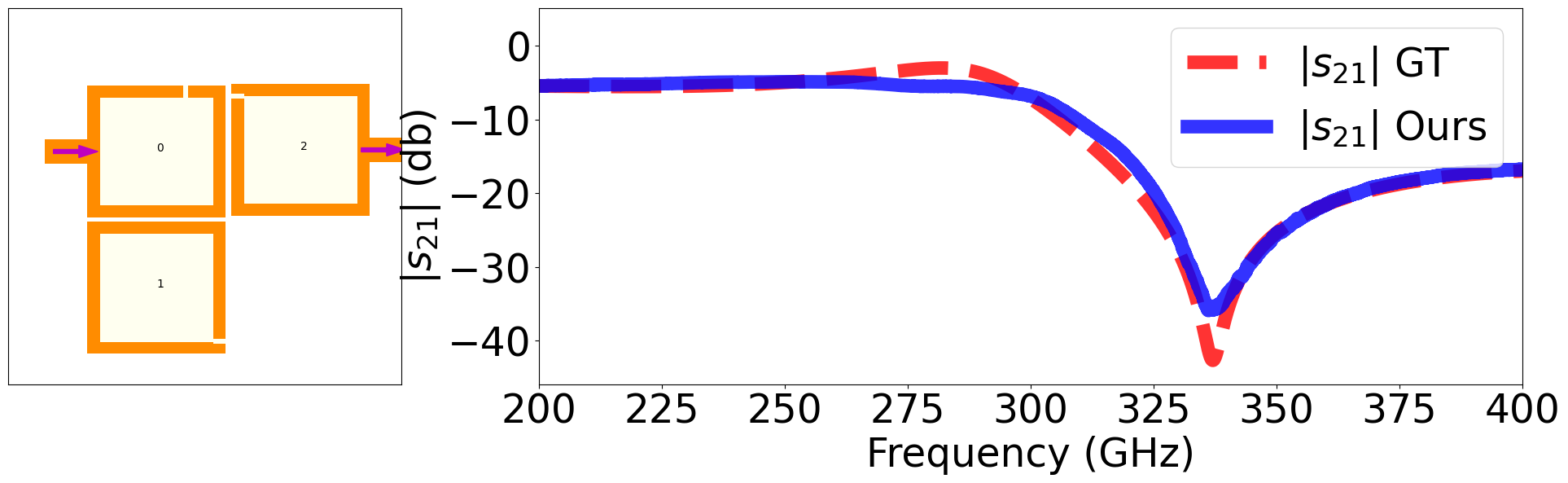}}
\end{minipage}%
\label{fig:4a}
}%
\subfigure[3-resonator circuit and its transfer function from \textbf{Circuit-GNN}]{
\begin{minipage}{0.5\linewidth}
\centering
\resizebox{\linewidth}{!}{\includegraphics[]{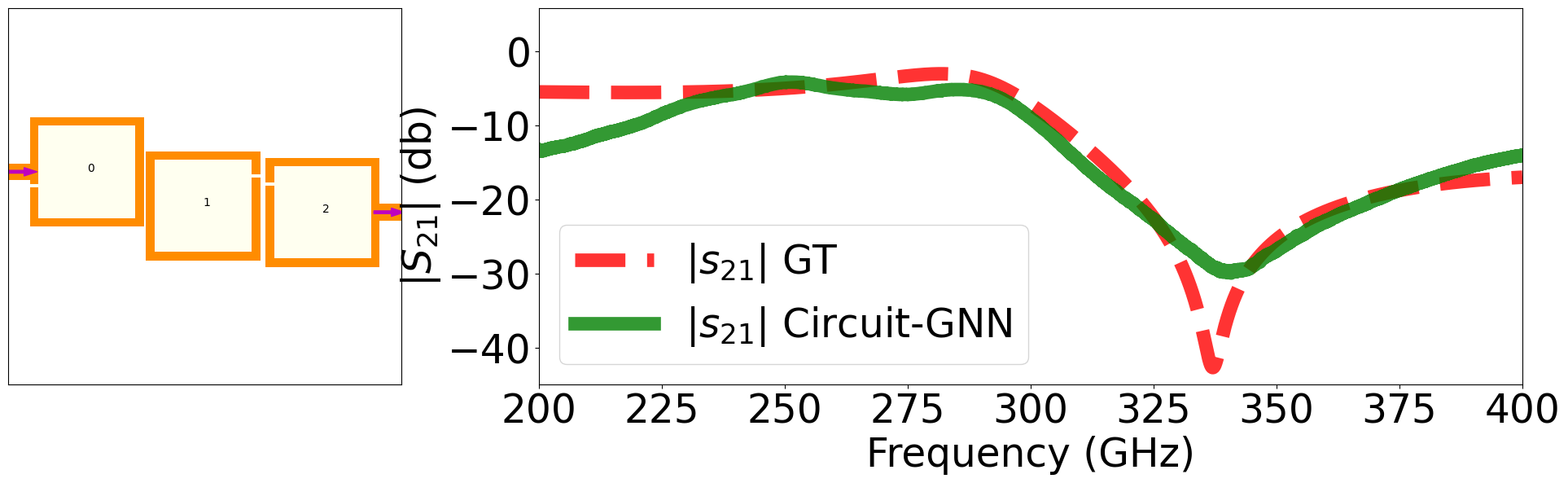}}
\end{minipage}%
\label{fig:4b}
}%

\subfigure[4-resonator circuit and its transfer function from \textbf{DCIDA}]{
\begin{minipage}{0.5\linewidth}
\centering
\resizebox{\linewidth}{!}{\includegraphics[]{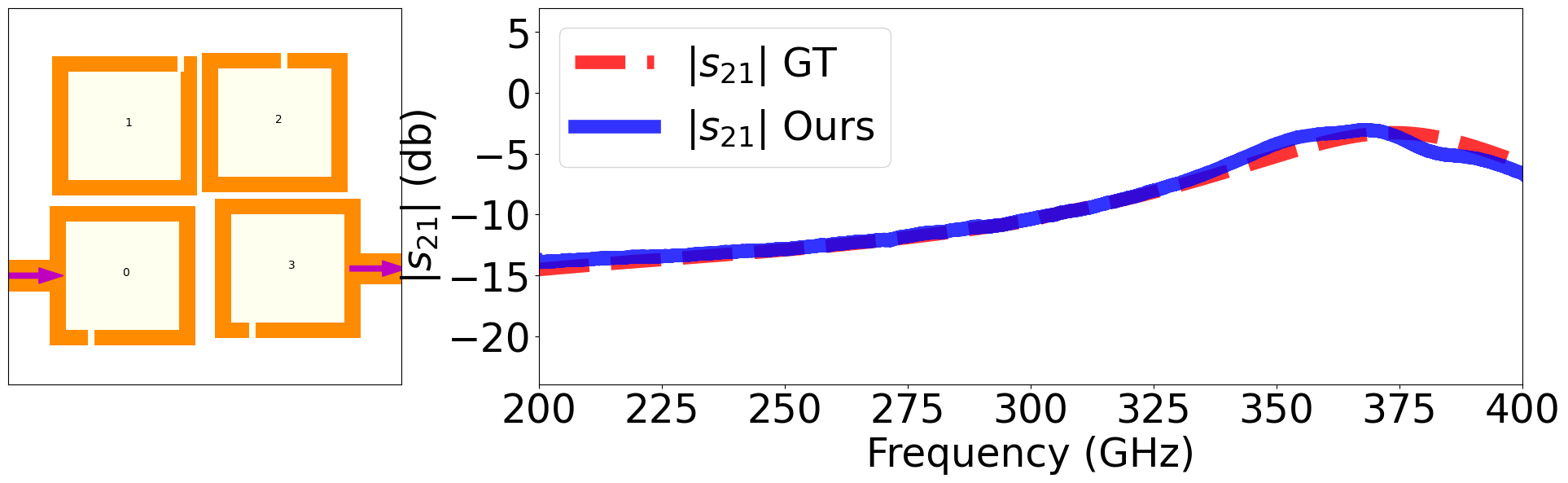}}
\end{minipage}%
\label{fig:4c}
}%
\subfigure[4-resonator circuit and its transfer function from \textbf{Circuit-GNN}]{
\begin{minipage}{0.5\linewidth}
\centering
\resizebox{\linewidth}{!}{\includegraphics[]{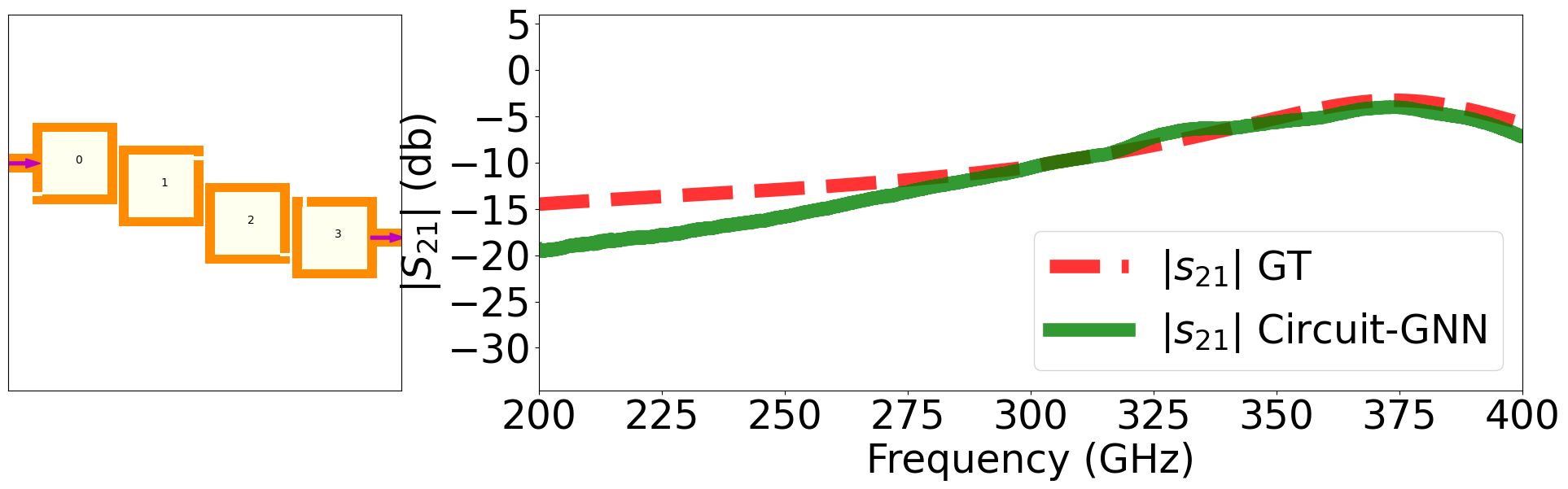}}
\end{minipage}%
\label{fig:4d}
}%

\subfigure[5-resonator circuit and its transfer function from \textbf{DCIDA}]{
\begin{minipage}{0.5\linewidth}
\centering
\resizebox{\linewidth}{!}{\includegraphics[]{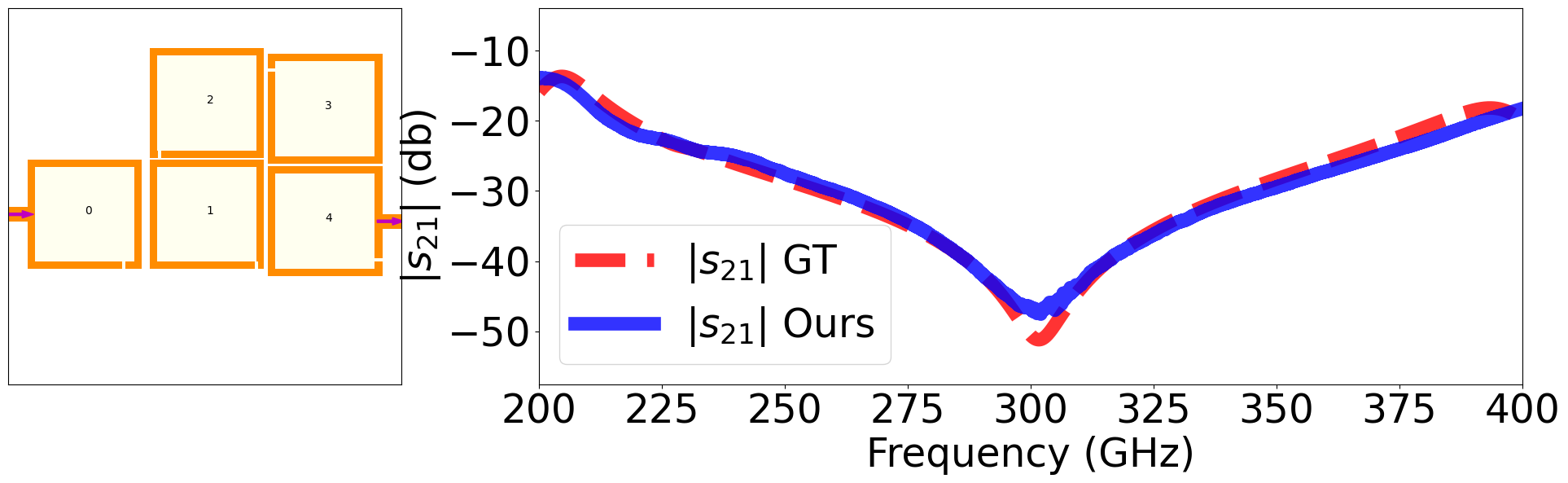}}
\end{minipage}%
\label{fig:4e}
}%
\subfigure[5-resonator circuit and its transfer function from \textbf{Circuit-GNN}]{
\begin{minipage}{0.5\linewidth}
\centering
\resizebox{\linewidth}{!}{\includegraphics[]{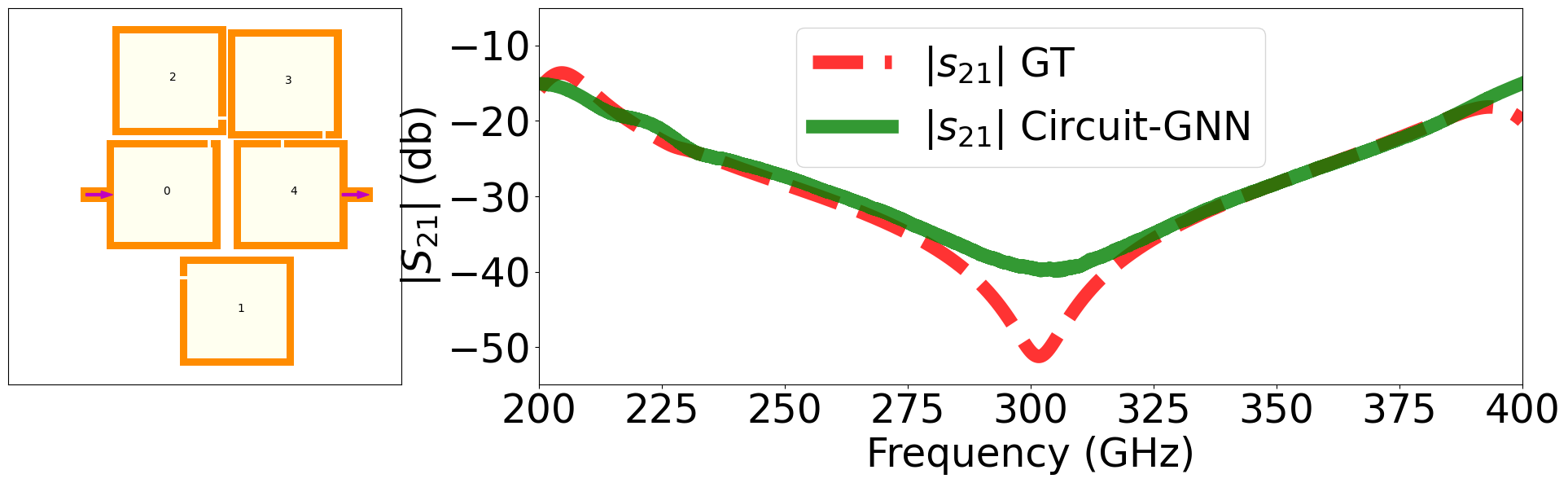}}
\end{minipage}%
\label{fig:4f}
}%

\subfigure[6-resonator circuit and its transfer function from \textbf{DCIDA}]{
\begin{minipage}{0.5\linewidth}
\centering
\resizebox{\linewidth}{!}{\includegraphics[]{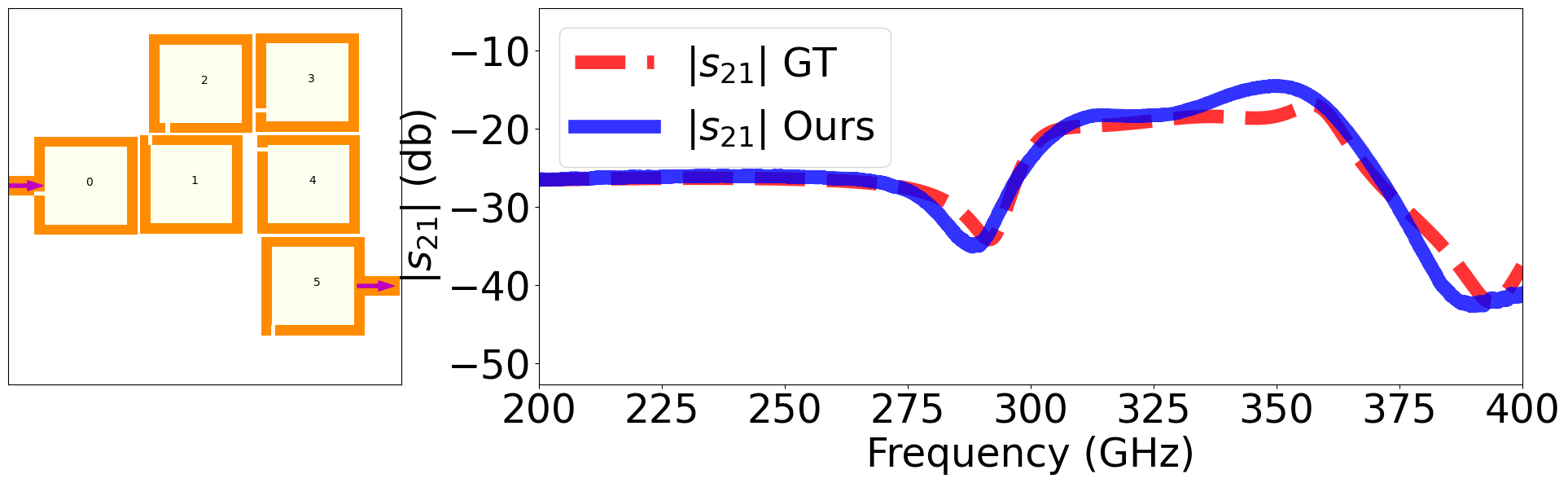}}
\end{minipage}%
\label{fig:4g}
}%
\subfigure[6-resonator circuit and its transfer function from \textbf{Circuit-GNN}]{
\begin{minipage}{0.5\linewidth}
\centering
\resizebox{\linewidth}{!}{\includegraphics[]{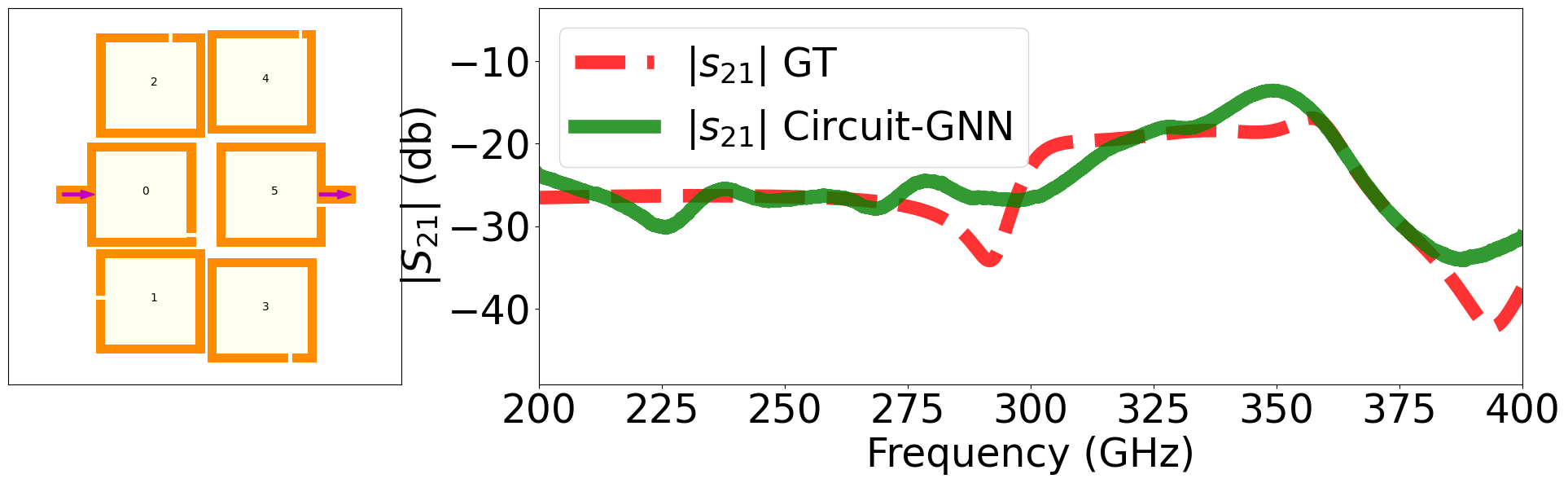}}
\end{minipage}%
\label{fig:4h}
}%
\centering
\vspace{-3mm}\caption{Visualization of inverse designs from DCIDA and Circuit-GNN with the \textcolor{red}{desirable transfer functions} (ground truth (GT)). Blue lines are \textcolor{blue}{transfer functions} of the generated circuits from Transformer-based DCIDA. Green lines are \textcolor{Green}{transfer functions} of the generated circuits from Circuit-GNN. More challenging cases are shown in Appendix \ref{Challenge}.}
\label{fig4}
\end{figure*}
In this section, we demonstrate the effectiveness of our proposed method through the following perspectives:
\begin{itemize}
    \item We compare the performance of DCIDA with the state-of-the-art method \textbf{Circuit-GNN} and \textbf{$\theta$-Resonance}\footnote{We define $\theta$-Resonance using Transformer to manage mixed actions as $\theta$-Resonance. We cannot reproduce RLDFCDO because the code is inaccessible.} in the inverse design of distributed circuits.
    \item We analyze the effect of interdependent mapping functions in DCIDA, the convergence of DCIDA and the hyper-parameter sensitivity in DCIDA. 
\end{itemize}

\subsection{Dataset and Experiment Setup}
\noindent \textbf{Dataset.} The circuits consist of $3, 4, 5$, and $6$ resonators in Circuit-GNN dataset\footnote{The Circuit-GNN dataset with harder examples are large size and balanced. Please see the website: \url{https://circuit-gnn.csail.mit.edu.}} have $4, 10, 9$ and $6$ topology types respectively. Templates of circuits differ in the topology type and number of resonators. We randomly sample $10$ transfer functions as target transfer functions from each template.

\noindent \textbf{Experiment Setup.} Circuit-GNN in inverse design utilizes the topology types as one of the inputs which are invisible to experts at the beginning of the inverse design. Therefore, we mask topology types for Circuit-GNN during the inverse design. To make the comparison fair, we applied the same pre-trained forward model as an approximator of the simulator to evaluate the transfer functions of all generated circuits. We assess the errors $\epsilon_\text{db}$ between the desired transfer function and the transfer functions derived from circuits generated by DCIDA, as well as those from Circuit-GNN and $\theta$-Resonance. A lower value of $\epsilon_\text{dB}$ indicates better generation performance. In the parameter settings, we apply the default settings in Circuit-GNN and $\theta$-Resonance. We denote DCIDA using a Transformer as Transformed-based DCIDA and define DCIDA with a multilayer perceptron (MLP) as MLP-based DCIDA. We keep the size of the two DCIDAs close to \qty{0.32}{\mega\byte} for a fair comparison. During the training of DCIDA, we use the ADAM optimizer with the learning rate $1 \times 10^{-5}$ and set iterations to $1500$, which is the same as $\theta$-Resonance does. In each iteration, we set the batch size to $1024$ with a mini-batch size of $512$ and the number of epochs to $1$. We set the renew rate $\alpha_{r}$ to $0.2$, the anomalous rate to $\alpha_{a}$ $0.2$, the $\beta_\text{KL}$ to $3$, the $\beta_{e}$ to $1$, the $\beta_\text{min}$ to $0.02$, and $\beta_\text{decay}$ to $0.993$. \vspace{-2mm}

\subsection{Comparison in Inverse Design}
\noindent \textbf{Without any assumption about topology types.} Table \ref{tab:performance1} and Table \ref{tab:performance2} show the comparison between Transformer-based DCIDA and baselines including Circuit-GNN and $\theta$-Resonance in the inverse design on Circuit-GNN dataset. $\theta$-Resonance performs worst, since the method insufficiently explores the designs in an infinite slate without any boundaries and interdependent mapping functions. Transformer-based DCIDA generates the 3-resonator circuits with the best performance, whose transfer functions more closely align with the target transfer functions. Notably, the average error of the transfer function of 3-resonator circuits is reduced by over $27\%$. In the generation of the 4-resonator circuits, although Transformer-based DCIDA decreases the average error by about $3\%$ and $11\%$ in topology type 2 and topology type 4 respectively, transformer-based DCIDA outperforms Circuit-GNN in most topology types with the average error dropping more than $20\%$. When generating 5-resonator circuits, Transformer-based DCIDA has comparable performance as the Circuit-GNN in the topology type 0, 1, 2, and 6, where the average error reduces less than $3\%$. However, in the remaining topology about 5-resonator circuits, the performance of generated circuits from Transformer-based DCIDA surpasses those from Circuit-GNN by reducing the average error by more than $7\%$. Transformer-based DCIDA also reduces the average errors of the 6-resonator circuits more than $10\%$. Figure~\ref{fig4} visualizes the superior capability of the Transformer-based DCIDA.

\noindent \textbf{Without any assumption about topology types and the number of resonators.} We randomly sample a target transfer function related to 5-resonator circuits. We mask its corresponding template information including topology type and the number of resonators. We compare the performance of generated circuits from Transformer-based DCIDA and Circuit-GNN only with the given transfer function. In Figure~\ref{fig5}, although both Circuit-GNN and DCIDA correctly figure out the target transfer function corresponding to a 5-resonator circuit, DCIDA generates the best 5-resonator circuits with the minimum $\epsilon_\text{db}$.

\subsection{Performance Analysis}
\begin{figure}{}
	\centering
	\subfigure[CDF of Pass-band IOU]{
		\begin{minipage}{0.46\linewidth}
			\includegraphics[width=1\linewidth]{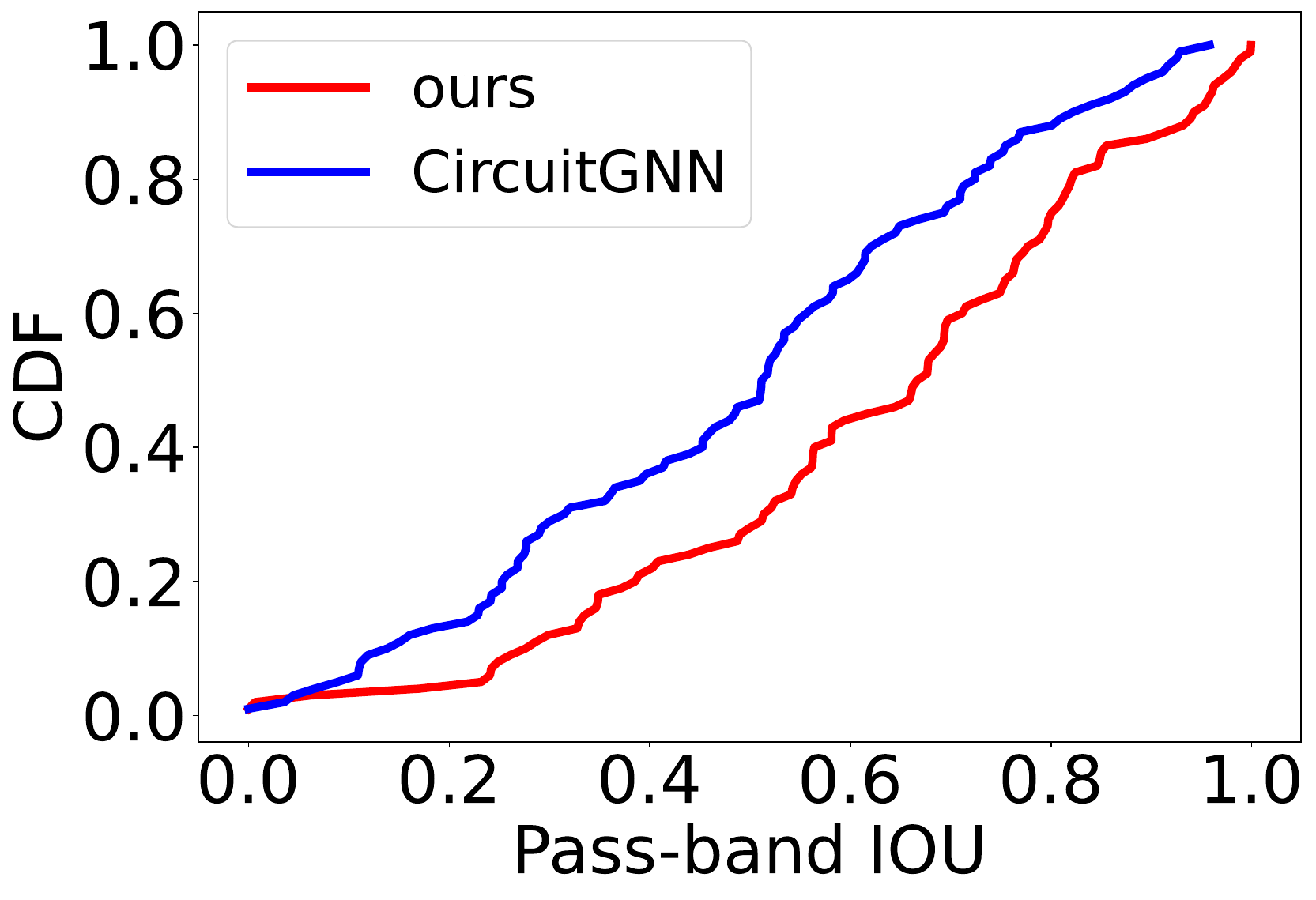} 
		\end{minipage}
		\label{fig:6a}
	}
    	\subfigure[CDF of Insertion Loss]{
    		\begin{minipage}{0.46\linewidth}
   		 	\includegraphics[width=1\linewidth]{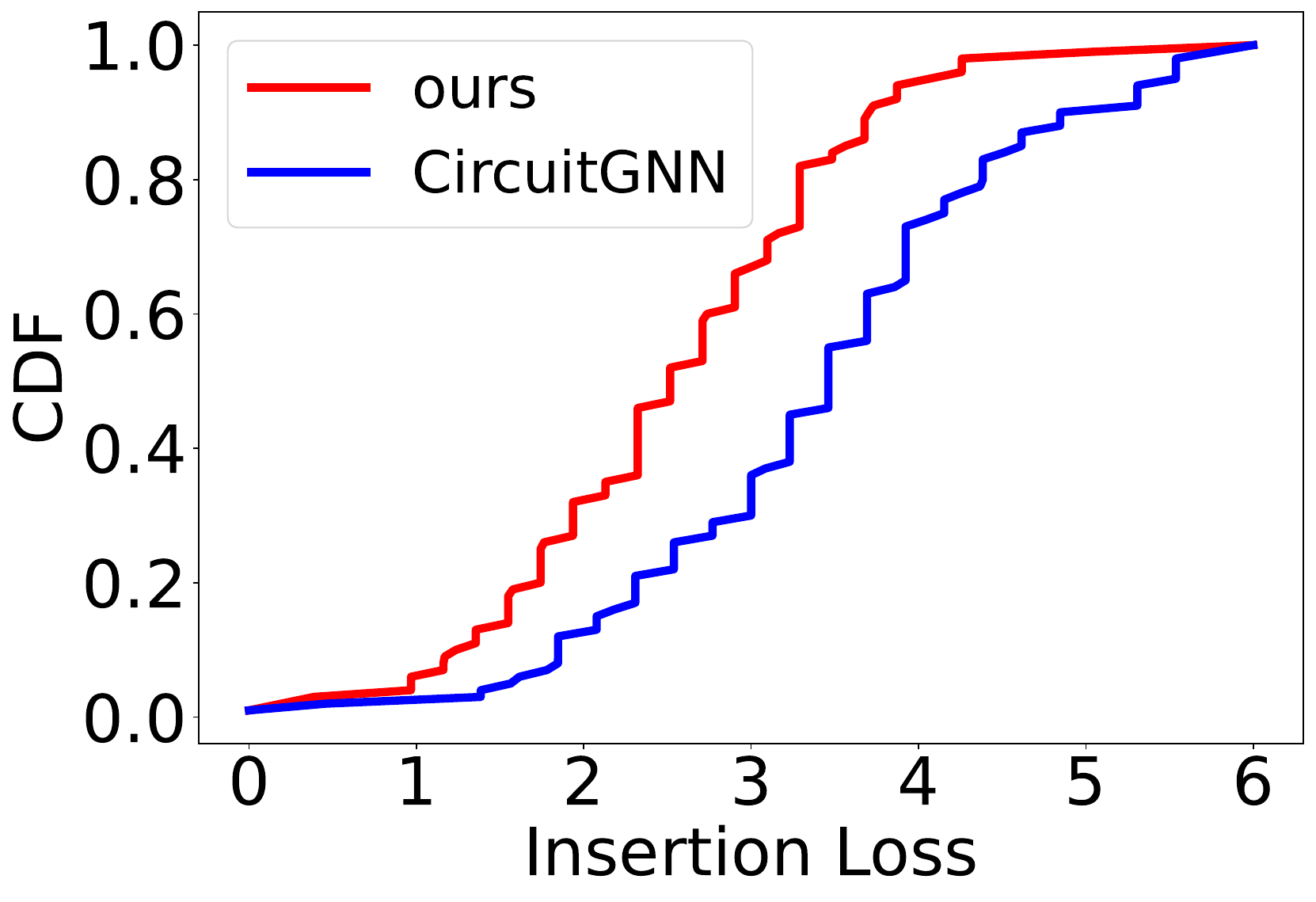}
    		\end{minipage}
		\label{fig:6b}
    	}
	\caption{The comparison of pass-band IOU and insertion loss between Transformer-based DCIDA and Circuit-GNN in inverse design of circuits.}
	\label{fig:6}
\end{figure}
 \noindent \textbf{Cumulative Distribution Function (CDF).} We sample $100$ generated circuits produced by DCIDA and Circuit-GNN, and utilize pass-band IOU metric and insertion loss~\cite{pmlr-v97-zhang19e} to evaluate the quality of the generated circuits. The pass-band IOU quantifies the proximity of the generated circuit's pass-band to the target band from a desired specification, while a smaller insertion loss indicates a better generated circuit which delivers more power in the desired bands. The Figure~\ref{fig:6} shows that generated circuits from DCIDA have better pass-band IOU and insertion loss, which has $0.63$ pass-band IOU and $2.55$db insertion loss while circuits from Circuit-GNN have $0.49$ pass-band IOU and $3.42$db insertion loss.
\begin{figure}[t]
    \centering
    \includegraphics[width=\columnwidth]{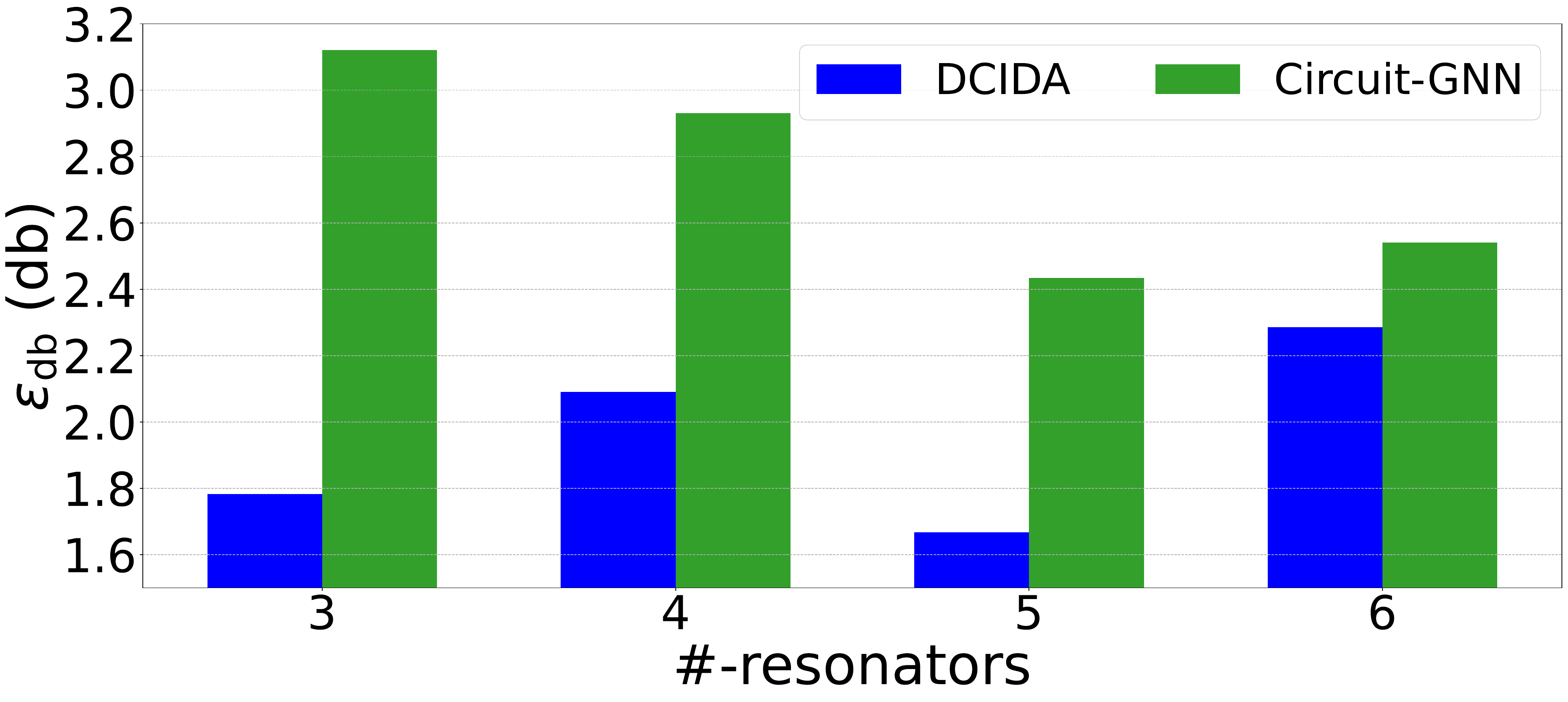}
    \caption{Error of inverse designs from Transformer-based DCIDA and Circuit-GNN without template information.}
    \label{fig5}
\end{figure}

\noindent \textbf{The effect of the boundaries and interdependent functions.} To further demonstrate the effectiveness of the boundaries and interdependent functions (IDF) in DCIDA, we conduct an ablation study to evaluate DCIDA (\textbf{w/ boundary + w/ IDF}) against the following two variants. Note that all variants of DCIDA are Transformer-based. \vspace{-2mm}
\begin{itemize}
    \item \textbf{w/ boundary + w/o IDF}: the variant of DCIDA maps actions $a^{(i)}_x$ and $a^{(i)}_y$ related to each resonator $i$ including the rightmost resonator into the position $x_{i}=a^{(i)}_x(B-a)$ and $y_{i}=a^{(i)}_y(B-a)+\frac{a-B}{2}$ within the boundary $\gB$.
    \item \textbf{w/o boundary + w/o IDF}: the variant of DCIDA has an extended boundary $\gB$ with the length of $B$ decided by $B=aQ + g(Q-1)$, where $Q\gg N$.
\end{itemize}\vspace{-2mm}
Figure \ref{fig8} shows average errors of inverse designs from the variants of DCIDA using different target transfer functions corresponding to different numbers of resonator circuits. The experimental results highlight the substantial contributions of the boundary and interdependent functions to the effectiveness of DCIDA. The boundary stands out as the most pivotal component, as evidenced by the rise in errors when it is absent. This degradation in performance occurs due to an expansion of the search space when the boundaries expands, and a limited number of training iterations is insufficient to explore the design space with such complexity. The interdependent functions further precisely map the actions related to resonators to locations within the boundaries by utilizing the relationship between resonators.\vspace{-2mm}

\begin{figure}
    \centering
    \includegraphics[width=\columnwidth]{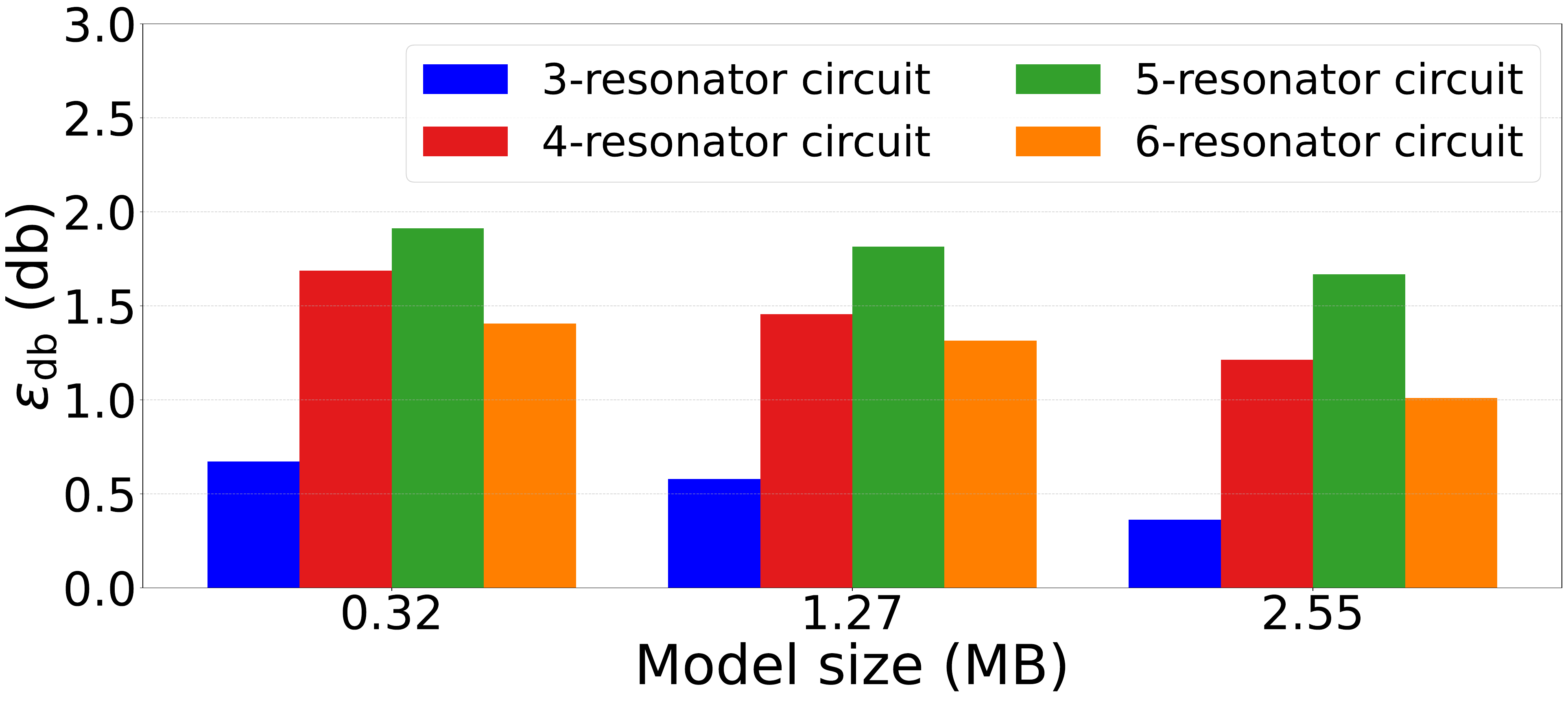}
    \caption{Error of inverse designs from Transformer-based DCIDA varies with different sizes of the Transformer model.}
    \label{fig7}
\end{figure}

\begin{figure}
    \centering
    \includegraphics[width=\columnwidth]{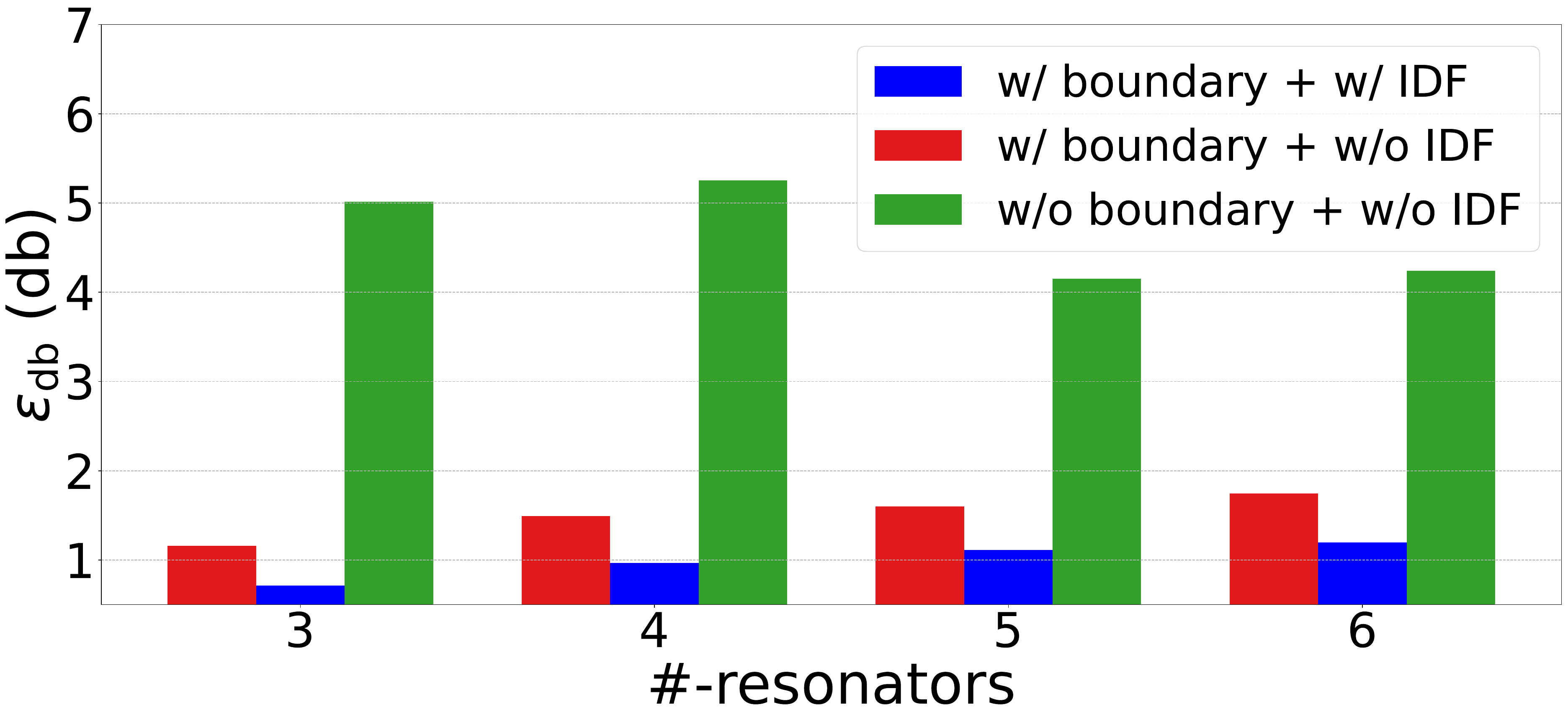}
    \caption{Average error of inverse designs from the variants of DCIDA on different circuits with 3, 4, 5, and 6 resonators.}
    \label{fig8}
\end{figure}

\begin{figure}
    \centering
    \includegraphics[width=\columnwidth]{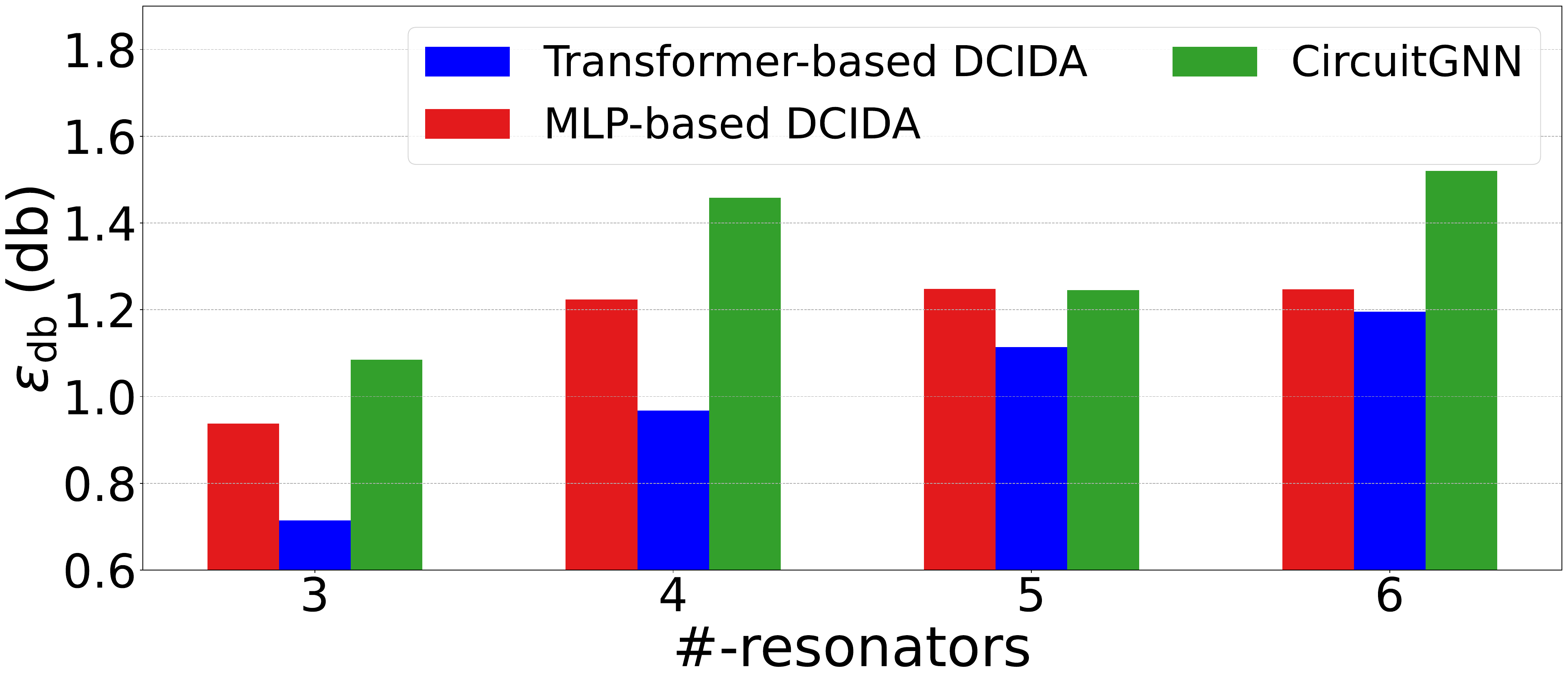}
    \caption{Average error of inverse designs from DCIDAs and Circuit-GNN on circuits with 3, 4, 5 and 6 resonators.}
    \label{fig9}
\end{figure}

\noindent \textbf{Hyper-parameter sensitivity} First, we investigate the effect of the size of the Transformer in DCIDA. We randomly sample target transfer functions from different number of resonator circuits. Figure~\ref{fig7} demonstrates that as the size of the Transformer increases, the error of inverse design from Transformer-based DCIDA decreases. Since the Transformer as the neural network in DCIDA produces $\pi_{\theta}$ as a collection of conditional probabilities for each dimension $i$ by decoding a constant tensor $\mI$, it is critical to have a neural network with good capacity for decoding, such that inter-dependencies can be learned. Second, to further show the importance of a neural network with good capacity, we compare the average error of inverse design from Transformer-based DCIDA, MLP-based DCIDA, and Circuit-GNN on different numbers of resonator circuits. For a fair comparison, we ensure that the models have the same size. Figure \ref{fig9} shows that Transformer-based DCIDA performs best. However, both of the variants of DCIDA generate better inverse designs than Circuit-GNN does. Due to limited space, the hyper-parameter sensitivity analysis related to the factors of objective $L(\pi (\theta))$ and the number of loops of backpropagation related to $\frac{ZE}{z}$ are given in Appendix \ref{para_sensitivity}.

\begin{figure}[t]
\centering
\subfigure[Original and $\alpha_{r}=0.8$]{
\begin{minipage}{0.48\linewidth}
\centering
\resizebox{\linewidth}{!}{\includegraphics[]{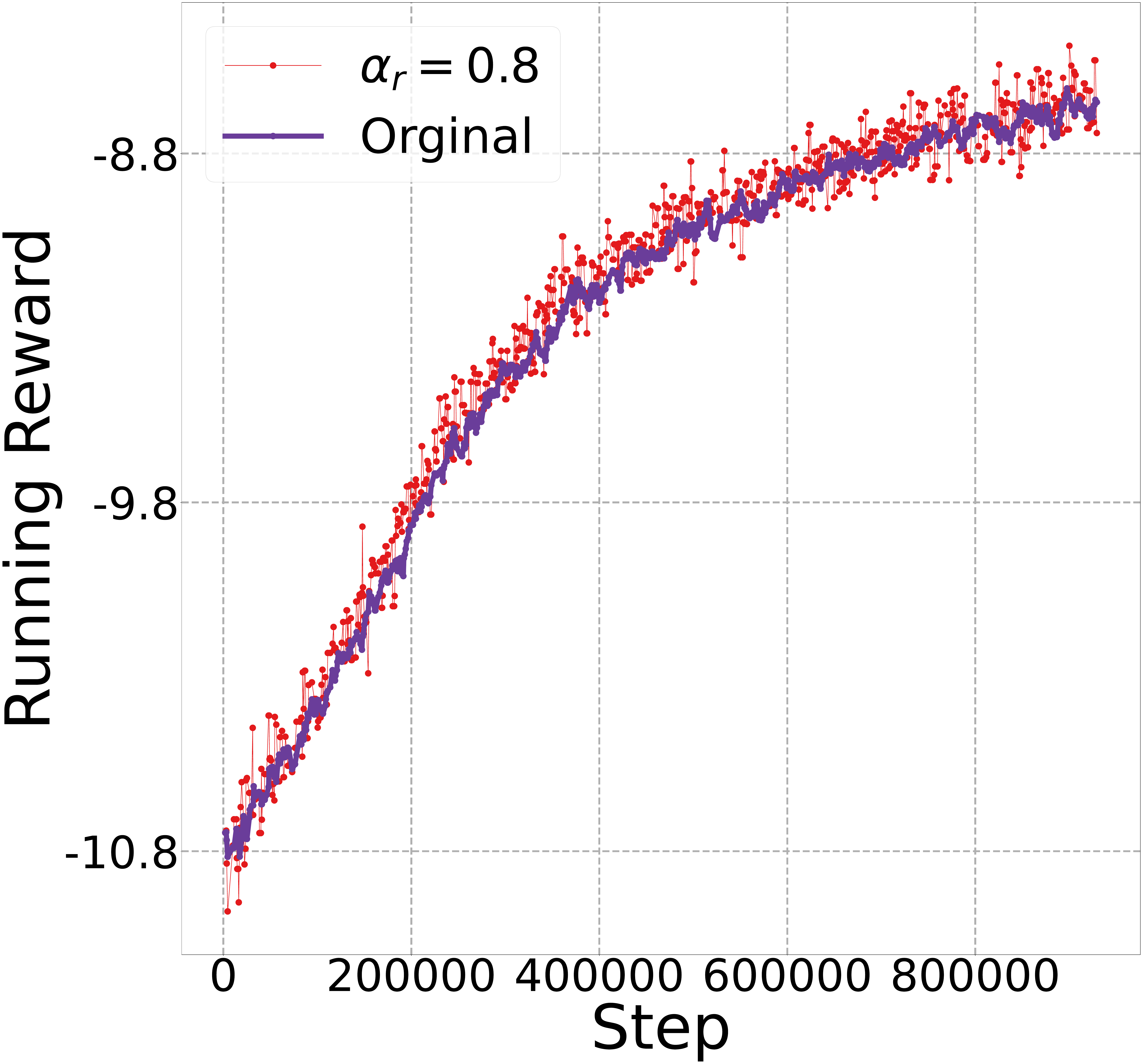}}
\end{minipage}%
\label{fig:10a}
}%
\subfigure[Original and $\beta_\text{KL}=0.3$]{
\begin{minipage}{0.48\linewidth}
\centering
\resizebox{\linewidth}{!}{\includegraphics[]{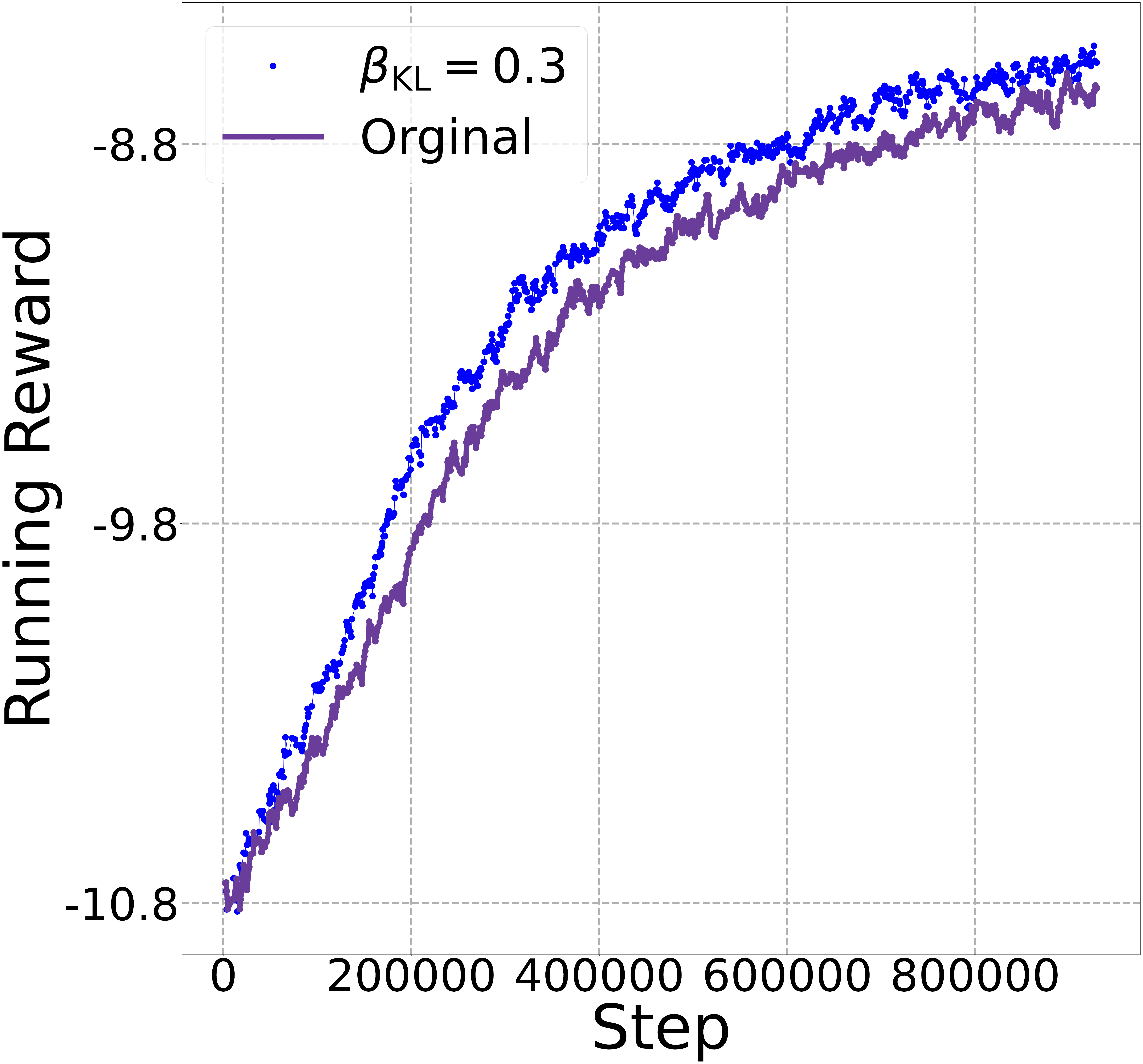}}
\end{minipage}%
\label{fig:10b}
}%

\subfigure[Original and linear $\beta_{e}$]{
\begin{minipage}{0.48\linewidth}
\centering
\resizebox{\linewidth}{!}{\includegraphics[]{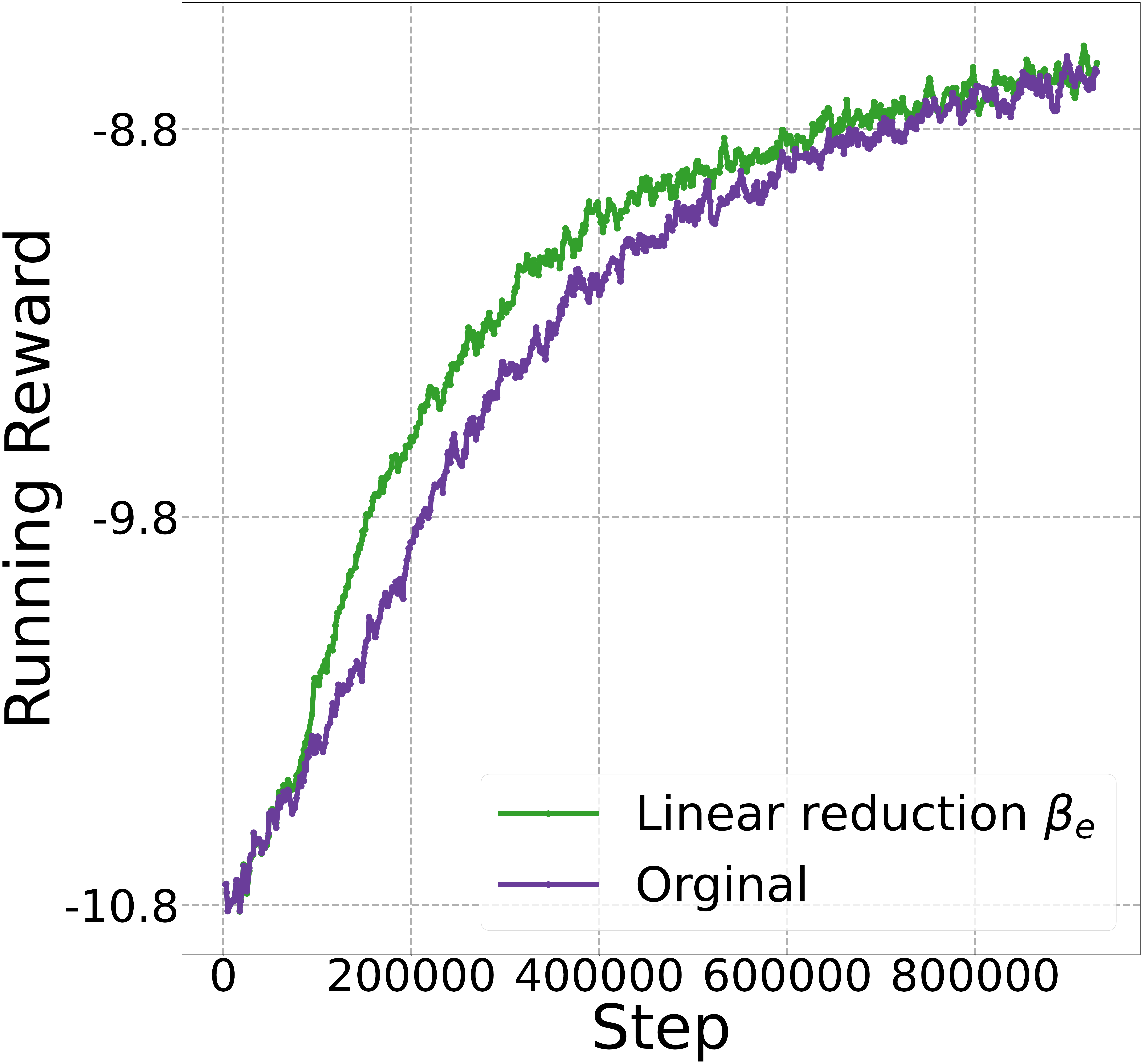}}
\end{minipage}%
\label{fig:10c}
}%
\subfigure[Linear $\beta_{e}$ and $\beta_\text{KL}=0.3$]{
\begin{minipage}{0.48\linewidth}
\centering
\resizebox{\linewidth}{!}{\includegraphics[]{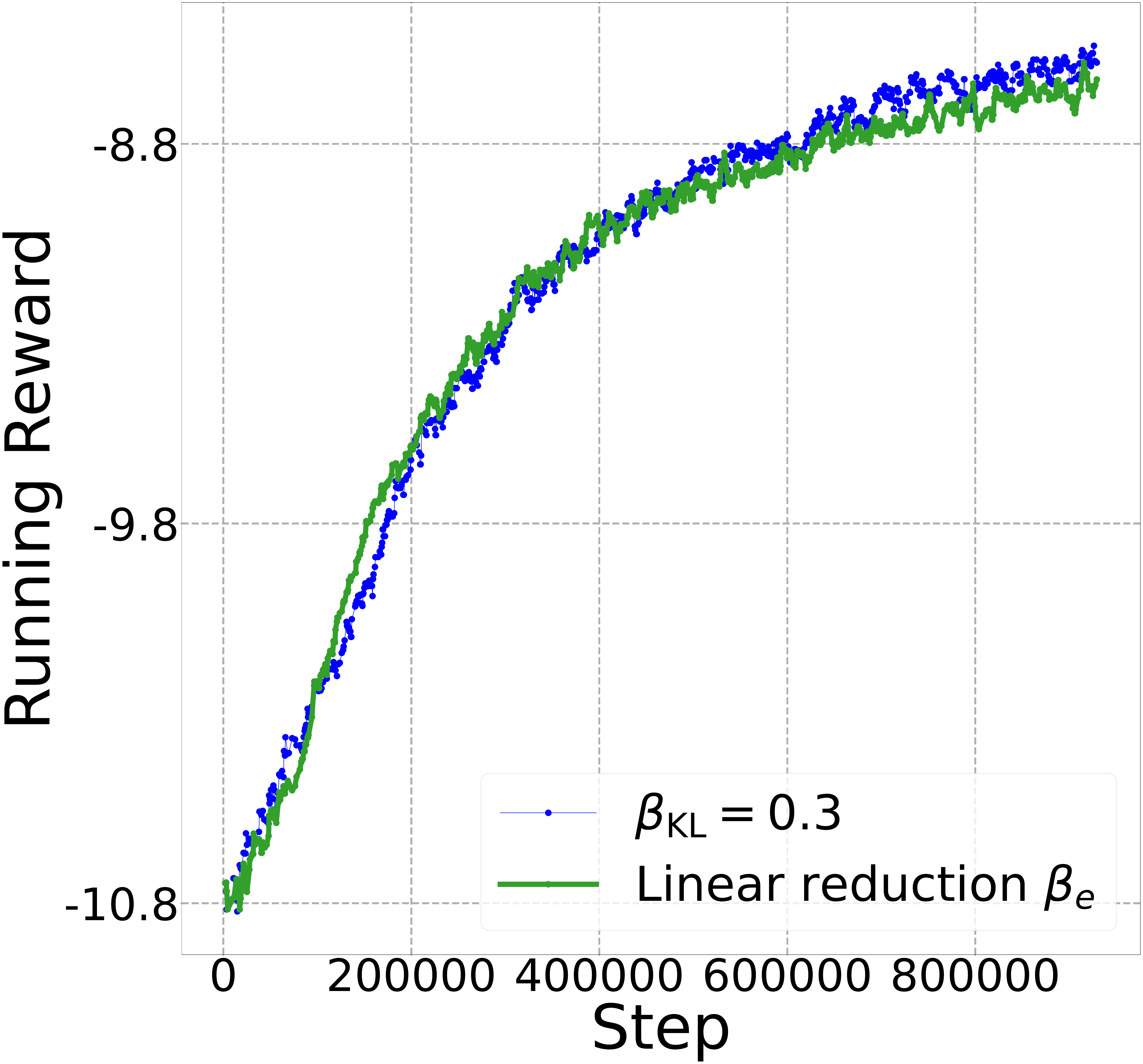}}
\end{minipage}%
\label{fig:10d}
}%
\centering
\vspace{-3mm}\caption{The plots show running reward during the training of Transformer-based DCIDA.} 
\label{fig10}
\end{figure}

\noindent \textbf{Convergence and running reward.} Figure \ref{fig10} shows the convergence of training Transformer-based DCIDA through running reward using different settings, including the original setting ($\alpha_{r}=0.2$, $\beta_\text{KL}=3$ and an exponential reduction $\beta_{e}$) with different $\alpha_{r}$, $\beta_\text{KL}$ and a linear reduction $\beta_{e}$. The linear reduction is $\beta_{e,(t)} = \beta_\text{min} + \frac{(T - t)}{T}(\beta_{e, (t-1)} - \beta_\text{min})$, where $T$ and $t$ are total iterations and the current iteration. Figure \ref{fig:10a} shows that when $\alpha_{r}=0.8$, the running rewards exhibit significant noise, indicating instability in the learning algorithm. Figure \ref{fig:10c} and \ref{fig:10d} with a linear reduction $\beta_{e}$ demonstrate that running rewards increase more significantly compared to those in the original settings. Figure \ref{fig:10b} shows that the trends in running rewards are similar when using the original setting with $\beta_\text{KL}=0.3$ and the default configuration. The experiments did not experience divergence issues, suggesting our method can train larger policy networks in a stable manner.\vspace{-2mm}

\section{Conclusion}
DCIDA is a design space exploration framework using single-step RL to produce near-optimal arrangements of distributed resonators that progressively mimic a target transfer function $s_{21}({\boldsymbol{\omega}})$. Unlike Circuit-GNN, DCIDA makes no prior assumptions regarding circuit topology types. For a given number of resonators, DCIDA trains a policy neural network that samples near-optimal ``raw'' design decisions and then maps these decisions injectively and interdependently to a physical representation. 
Experiments on the Circuit-GNN dataset demonstrate the superior performance of DCIDA. Compared to state-of-the-art methods, such as Circuit-GNN and $\theta$-Resonance, DCIDA generates distributed circuits that more accurately align with their target transfer functions.

\nocite{langley00}

\bibliography{paper}
\bibliographystyle{icml2023}

\newpage
\appendix
\onecolumn
\section{Proof of Theorems}
\label{theorems_proof}
\subsection{Proof of Theorem \ref{th1}}
\begin{proof}
Given a distributed circuit consist of $N$ square resonators with the lengths equal to $a$, we can easily compute the maximum length of a distributed circuit with the $N$ resonators as
\begin{equation*}
    B=aN + g(N-1).
\end{equation*} With the maximum length $B$, when defining the center of the leftmost resonator on $x$-axis can be $0$, the center of the rightmost resonator on $x$-axis can be $B-a$. Meanwhile, when setting the center of the uppermost resonator on $y$-axis as $\frac{B-a}{2}$, the center of the lowermost resonator on $y$-axis can be $\frac{a-B}{2}$. Finally, we summarize that the locations of centers of all resonators bounded by the area 
\begin{equation*}
    \gP=\{(x,y)|x\in [0,B-a], y\in[\frac{a-B}{2}, \frac{B-a}{2}]\}.
\end{equation*} As the lengths of resonators in a distributed circuit are $a$, all square resonators can be bounded within the boundary $\gB \subseteq \gM$ and $\gP \subseteq \gB$, where $\gM$ is the circuit space and 
\begin{equation*}
    \gB=\{(x,y)|x\in [-\frac{a}{2}, \frac{2B-a}{2}], y\in [-\frac{B}{2}, \frac{B}{2}]\}.
\end{equation*}
\end{proof}

\subsection{Proof of Theorem \ref{th2}}
\begin{proof}
The action $a^{(i)}_{x}$ and $a^{(i)}_{y}$ sampled from beta distributions range from $[0, 1]$ in the action space $\gA$. With the $a^{(i)}_{d}$ and the center $(x_{i-1}, y_{i-1})$ of a last resonator, we can map the actions $a^{(i)}_{x}$ and $a^{(i)}_{y}$ from action space $\gA$ to the area $\gP \subseteq \gM$. Therefore, the location of the center of a current resonator can be determined by the mapping. Based on the Theorem \ref{th1}, all resonators should be bounded within the area $\gB$. Note that we have the default center $(0,0)$ of the leftmost (first) resonator, then we sequentially add the remaining resonators from left to right based on a last resonator.

\textbf{A current resonator is added \underline{above} the last resonator when $a^{(i)}_{d} = 0$.} The center of the current resonator can be $x_{i-1} \leq x_{i} \leq \min(x_{i-1}+d_s, B-2a-d_g)$ and $y_{i} = \min(y_{i-1}+a+d_g, \frac{B-a}{2})$. For solving the $x_{i}$, we define a transformation function depending on the center of the last resonator $h^{(i)}_{x}: \beta^{(i)}_{x}a^{(i)}_{x}+ \gamma^{(i)}_{x} \rightarrow x_{i}$. With the range of $a^{(i)}_{x}$ from $[0,1]$, we can solve $x_{i} = a^{(i)}_{x}\min(x_{i-1}+d_s, B-2a-d_g)+(1-a^{(i)}_{x})x_{i-1}$, where the $\beta^{(i)}_{x}$ and $\gamma^{(i)}_{x}$ can be
\begin{equation}
\begin{split}
\label{appendix_eq1}
    \sigma^{(i)}_{x} &= x_{i-1}, \\
    \beta^{(i)}_{x} &= \min(x_{i-1}+d_s, B-2a-d_g) - x_{i-1}.
\end{split}
\end{equation}

\textbf{A current resonator is added \underline{below} the last resonator when $a^{(i)}_{d} = 1$.} $x_{i}$ in the center of the current resonator still have the range $x_{i-1} \leq x_{i} \leq \min(x_{i-1}+d_s, B-2a-d_g)$ but $y_{i} = min(y_{i-1}-a-d_{g}, \frac{a-B}{2})$. With the range of $a^{(i)}_{x}$ from $[0,1]$, the center can be computed as $x_{i} =a^{(i)}_{x}\min(x_{i-1}+d_s, B-2a-d_g)+(1-a^{(i)}_{x})x_{i-1}$, where $\beta^{(i)}_{x}$ and $\sigma^{(i)}_{x}$ are the same as the Eq.(\ref{appendix_eq1}).

\textbf{A current resonator is added to the \underline{right} of the last resonator when $a^{(i)}_{d} = 2$.} The center of the current resonator can be $x_{i} = \min(x_{i-1}+a+d_g, B-2a-d_g)$ with $\max(y_{i-1}-d_s, \frac{a-B}{2}) \leq y_{i} \leq \max(y_{i-1}+d_s, \frac{B-a}{2})$. Note that $a^{(i)}_{y}$ ranges from $0$ to $1$. When applying a transformation functions $h^{(i)}_{y}: \beta^{(i)}_{y} a^{(i)}_{y} + \sigma^{(i)}_{y} \rightarrow y_{i}$ depending on the center of the last resonator, we solve the $y_{i} = a^{(i)}_{y}\min(y_{i-1}+d_s, \frac{B-a}{2}) + (1 - a^{(i)}_{y})\max(y_{i-1}-d_s, \frac{a-B}{2})$, where we have $\beta^{(i)}_{y}$ and $\sigma^{(i)}_{y}$ as
\begin{align*}
    \sigma^{(i)}_{y} &= \min(y_{i-1}-d_s, \frac{a-B}{2})\\ 
    \beta^{(i)}_{y} &= \max(y_{i-1}+d_s, \frac{B-a}{2}) -\min(y_{i-1}-d_s, \frac{a-B}{2})
\end{align*}
Finally, we summarize the interdependent functions based on the $a^{(i)}_{d}$,
\begin{align*}
h^{(i)}_{x}&=
\begin{cases}
a^{(i)}_{x}\min(x_{i-1}+d_s, B-2a-d_g)+(1-a^{(i)}_{x})x_{i-1}, \qquad \qquad \quad & \text{$a^{(i)}_{d} = 0$}\\
a^{(i)}_{x}\min(x_{i-1}+d_s, B-2a-d_g)+(1-a^{(i)}_{x})x_{i-1}, \qquad \qquad \quad &\text{$a^{(i)}_{d} = 1$}\\
\min(x_{i-1}+a+d_g, B-2a-d_g), \qquad \qquad \qquad  \qquad \qquad \quad & \text{$a^{(i)}_{d} = 2$}\\
\end{cases}\\
h^{(i)}_{y}&=
\begin{cases}
\min(y_{i-1}+a+d_g, \frac{B-a}{2}), &\text{$a^{(i)}_{d} = 0$}\\
\max(y_{i-1}-a-d_g, \frac{a-B}{2}), &\text{$a^{(i)}_{d} = 1$}\\
a^{(i)}_{y}\min(y_{i-1}+d_s, \frac{B-a}{2}) + (1 - a^{(i)}_{y})\max(y_{i-1}-d_s, \frac{a-B}{2}), &\text{$a^{(i)}_{d} = 2$}\\
\end{cases}
\end{align*}
\end{proof}

\subsection{Proof of Theorem \ref{th3}}
\begin{proof}
In the action space $\gA$, the actions $a^{(n)}_{x}$ and $a^{(n)}_{y}$ from beta distributions with $[0, 1]$ relates to the rightmost resonator $n$. Note that we denote the rightmost (final) resonator as resonator $n$ while any remaining resonators as resonator $i$, where $i < n$. Based on the Theorem \ref{th1}, centers of resonators should be bounded within the area $\gP \subseteq \gB$, and all resonators with the length $a$ should be inside the boundary $\gB \subseteq \gM$. 

Therefore, \textbf{a current resonator can be assigned \underline{above} the last resonator when $a^{(n)}_{d} = 0$.} The center of a current resonator $(x_{n}, y_{n})$ should be $x_{n-1}\leq x_{n} \leq \min(x_{n-1}+d_s, B-a)$ and $y_{n} = min(y_{n-1}+a+d_{g}, \frac{B-a}{2})$. For transforming the actions $a^{(n)}_{x}$ into the circuit space $\gM$, we design a transformation function as $h^{(n)}_{x}: \beta^{(n)}_xa^{(n)}_{x} + \sigma^{(n)}_x \rightarrow x_{n}$ to obtain $x_{n} =a^{(n)}_{x}\min(x_{n-1}+d_s, B-a)+(1-a^{(n)}_{x})x_{n-1}$, where $\sigma^{(n)}_x$ and $\beta^{(n)}_x$ are respectively equal to
\begin{equation}
    \begin{split}
    \label{eq26}
    \sigma^{(n)}_x &= x_{n-1},\\
    \beta^{(n)}_x &= \min(x_{n-1}+d_s, B-a) - x_{n-1}.
    \end{split}
\end{equation}

Similarly, \textbf{a current resonator can be added \underline{below} the last resonator when $a^{(n)}_{d} = 1$}. The center of the current resonator should satisfy $x_{n-1}\leq x_{n} \leq \min(x_{n-1}+d_s, B-a)$ but $y_{n} = min(y_{n-1}-a-d_{g}, \frac{a-B}{2})$. With the same $\beta^{(n)}_{x}$ and $\sigma^{(n)}_{x}$ as the Eq.(\ref{eq26}), we solve $x_{n} =a^{(n)}_{x}\min(x_{n-1}+d_s, B-a)+(1-a^{(n)}_{x})x_{n-1}$.  

Finally, \textbf{a current resonator can be added to the \underline{right} of the last resonator when $a^{(n)}_{d}=2$.} The center of the current resonator can be $x_n = \min(x_{n-1}+a+d_g, B-a)$ with $\max(y_{n-1}-d_s, \frac{a-B}{2}) \leq y_n \leq \max(y_{n-1}+d_s, \frac{B-a}{2})$. When applying a transformation function $h^{(n)}_{y}: \beta^{(n)}_{y} a^{(n)}_{y} + \sigma^{(n)}_{y} \rightarrow y_{n}$, we obtain the $y_{n} =a^{(n)}_{y}\min(y_{n-1}+d_s, \frac{B-a}{2}) + (1 - a^{(n)}_{y})\max(y_{n-1}-d_s, \frac{a-B}{2})$, where we have $\sigma^{(n)}_{y}$ and $\beta^{(n)}_{y}$ as
\begin{equation*}
    \begin{split}
    \sigma^{(n)}_{y} &= \max(y_{i-1}-d_s, \frac{a-B}{2}),\\ 
    \beta^{(n)}_{y} &= \min(y_{i-1}+d_s, \frac{B-a}{2}) -\max(y_{i-1}-d_s, \frac{a-B}{2})
    \end{split}
\end{equation*}
Finally, we summarize the interdependent functions based on the $a^{(n)}_{d}$,
\begin{align*}
h^{(n)}_{x}&=
\begin{cases}
a^{(n)}_{x}\min(x_{n-1}+d_s, B-a)+(1-a^{(n)}_{x})x_{n-1},  \quad \quad \quad \ \quad \ \quad \quad \quad &\text{$a^{(n)}_{d} = 0$}\\
a^{(n)}_{x}\min(x_{n-1}+d_s, B-a)+(1-a^{(n)}_{x})x_{n-1},   \quad \quad \quad \ \quad \ \quad \quad \quad &\text{$a^{(n)}_{d} = 1$}\\
\min(x_{n-1}+d_g+a, B-a), \qquad \qquad \qquad \qquad \qquad \quad& \text{$a^{(n)}_{d} = 2$}\\
\end{cases}\\
h^{(n)}_{y}&=
\begin{cases}
\min(y_{n-1}+a+d_g, \frac{B-a}{2}), &\text{$a^{(n)}_{d} = 0$}\\
\max(y_{n-1}-a-d_g, \frac{a-B}{2}) &\text{$a^{(n)}_{d} = 1$}\\
a^{(n)}_{y}\min(y_{n-1}+d_s, \frac{B-a}{2})+(1 - a^{(n)}_{y})\max(y_{n-1}-d_s, \frac{a-B}{2}), &\text{$a^{(n)}_{d} = 2$}\\
\end{cases}
\end{align*}
\end{proof}

\section{Case Study about Mapping Actions to Designs} 
\label{case_study1}
\begin{figure*}[t]
    \centering
\includegraphics[width=\linewidth]{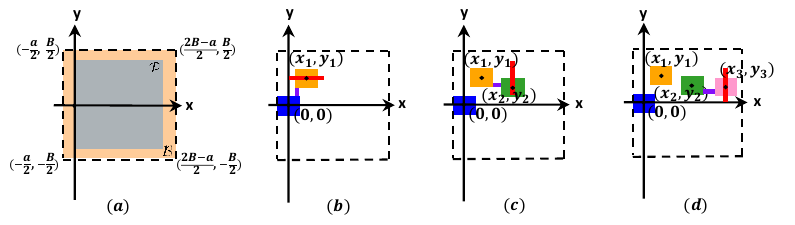}
    \caption{An example demonstrates a procedure in which DCIDA sequentially places all resonators within the boundary $\gB$.}
    \label{fig11}
\end{figure*}

Figure \ref{fig11} shows that DCIDA follows the rules of Theorem \ref{th1}, Theorem \ref{th2} and Theorem \ref{th3} to map actions into the boundary $\gB$ during the generation of a 4-resonator circuit.  In the Figure \ref{fig11}, \textcolor{orange}{\textbf{Orange area}} and \textcolor{gray}{\textbf{gray area}} are the boundaries $\gB$ and $\gP$, shown in (a). We map all resonators in the boundary by the following procedure:
\begin{itemize}
    \item \textbf{The location of \textcolor{blue}{the leftmost (first) resonator}.} In the Figure \ref{fig11} (b), we define a default position $(0,0)$ as the center $(x_{0}, y_{0})$ of the leftmost (first) resonator.
    \item \textbf{The location of \textcolor{Orange}{the second resonator}.} With the position $(0,0)$ of the first resonator and actions $a^{(1)}_x$ and $a^{(1)}_y$ sampled from distributions $\gF$, we follow the rule of Theorem \ref{th2} to map the actions into the position $(x_1, y_1)$ as the center of the second resonator, shown in the Figure \ref{fig11} (b). 
    \item \textbf{The location of \textcolor{ForestGreen}{the third resonator.}} Figure \ref{fig11} (c) illustrates that with the help of the second resonator's center $(x_1, y_1)$ and the interdependent mapping functions based on Theorem \ref{th2}, we map the actions $a^{(2)}_x$ and $a^{(2)}_y$ into the position $(x_2, y_2)$ as the center of the third resonator.
    \item \textbf{The location of \textcolor{pink}{the rightmost (final) resonator.}} With the center $(x_2, y_2)$ of the third resonator, we apply the interdependent functions from Theorem \ref{th3} to map actions $a^{(3)}_x$ and $a^{(3)}_y$ into position $(x_3, y_3)$ as the center of the rightmost (final) resonator, as illustrated in Figure \ref{fig11}(d).
\end{itemize}
Note that the \textcolor{Plum}{\textbf{gap}} and \textcolor{Red}{shift} are related to $d_g$ and $d_s$ from actions $a_f$, $a_{us}$ and $a_{ug}$. The computation of $d_g$ and $d_s$ are shown in the section \ref{mapping}.

\section{Performance on Challenge Cases}
\label{Challenge}
There exist some complex transfer functions $s_{21}$. These functions include multiple peaks, valleys and kinks. Generating distributed circuits to meet the complex transfer functions can be an challenging task in the inverse design of distributed circuits. In this section, we visualize the inverse designs generated by Transformer-based DCIDA and Circuit-GNN, which aim to satisfy complex transfer functions. In terms of the visualization, we compare the capability of Transformer-based DCIDA with that of Circuit-GNN in solving challenging tasks in the inverse design of distributed circuits. 

\begin{figure*}[t]
\centering
\subfigure[3-resonator circuit and its transfer function from \textbf{DCIDA}]{
\begin{minipage}{0.5\linewidth}
\centering
\resizebox{\linewidth}{!}{\includegraphics[]{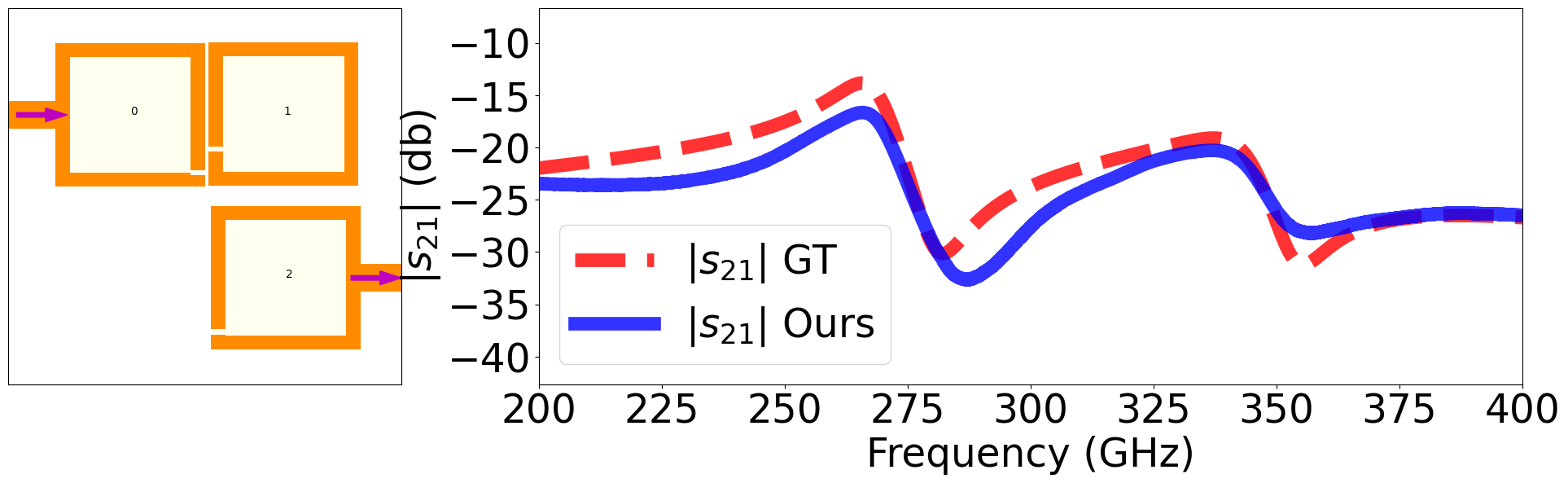}}
\end{minipage}%
\label{fig:12a}
}%
\subfigure[3-resonator circuit and its transfer function from \textbf{Circuit-GNN}]{
\begin{minipage}{0.5\linewidth}
\centering
\resizebox{\linewidth}{!}{\includegraphics[]{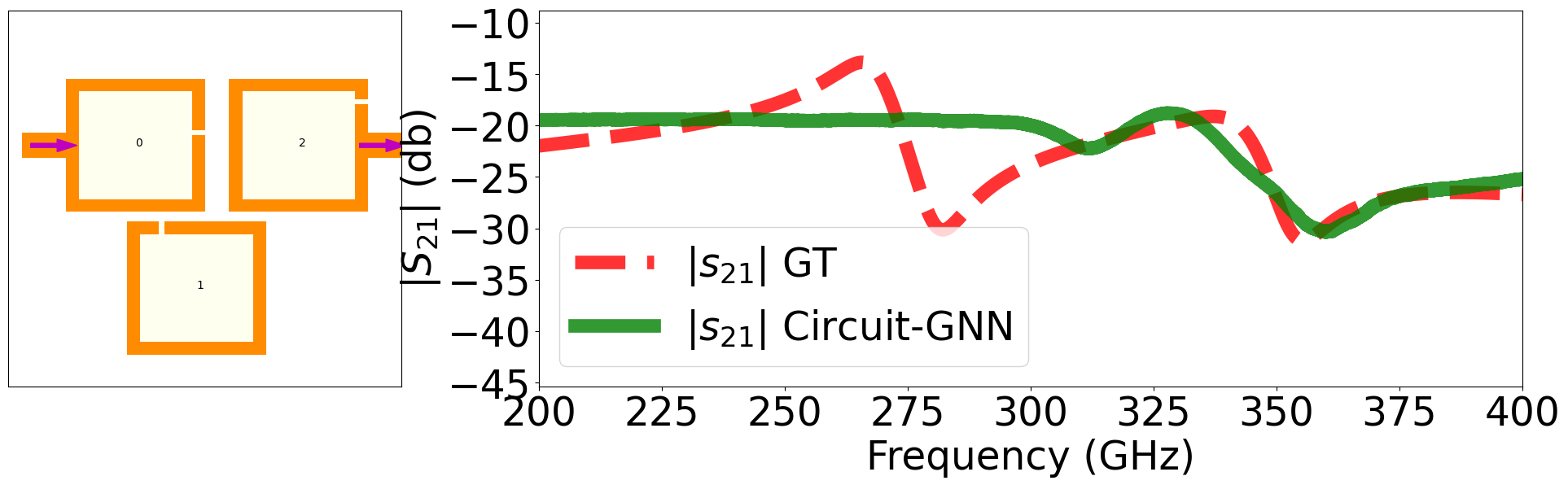}}
\end{minipage}%
\label{fig:12b}
}%

\subfigure[4-resonator circuit and its transfer function from \textbf{DCIDA}]{
\begin{minipage}{0.5\linewidth}
\centering
\resizebox{\linewidth}{!}{\includegraphics[]{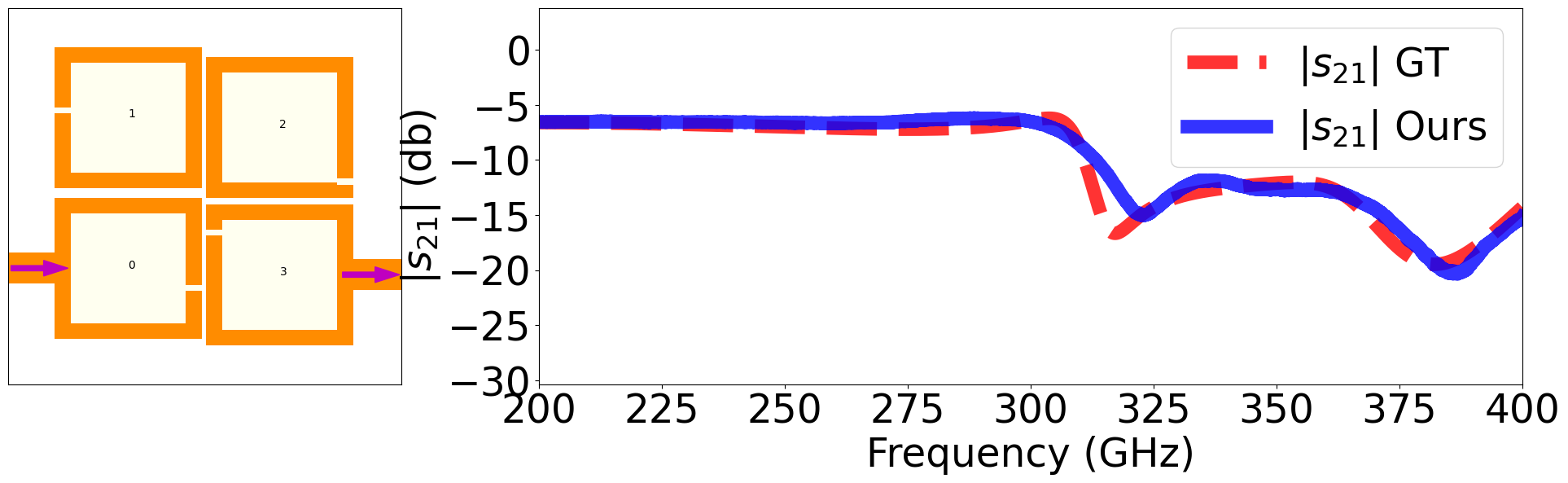}}
\end{minipage}%
\label{fig:12c}
}%
\subfigure[4-resonator circuit and its transfer function from \textbf{Circuit-GNN}]{
\begin{minipage}{0.5\linewidth}
\centering
\resizebox{\linewidth}{!}{\includegraphics[]{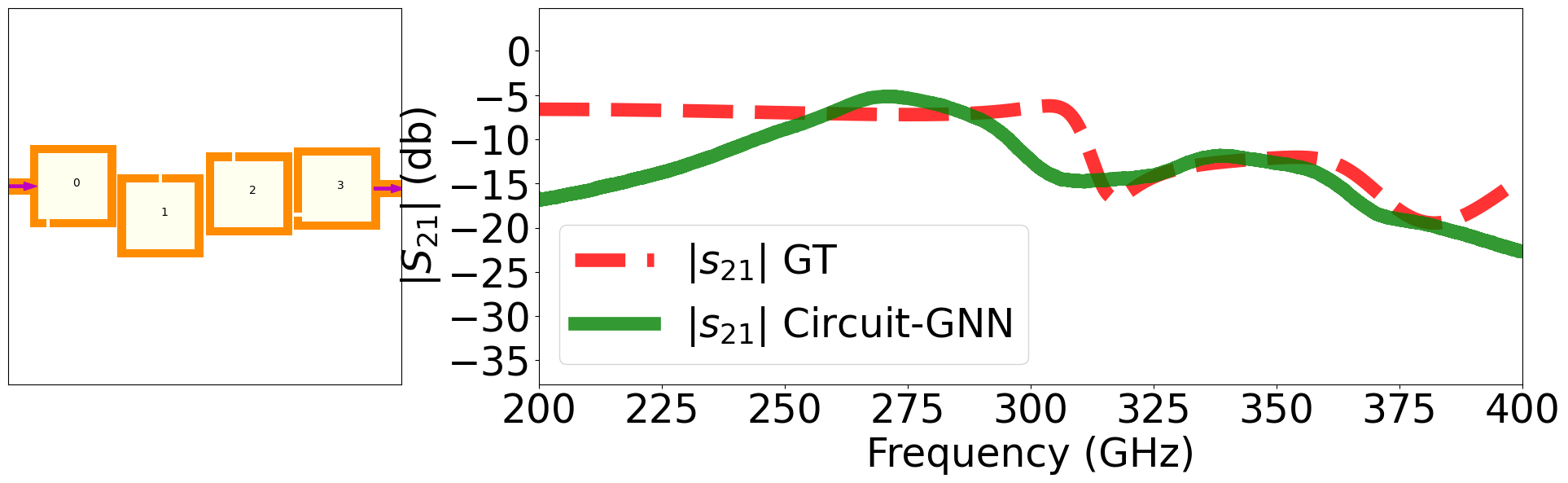}}
\end{minipage}%
\label{fig:12d}
}%

\subfigure[5-resonator circuit and its transfer function from \textbf{DCIDA}]{
\begin{minipage}{0.5\linewidth}
\centering
\resizebox{\linewidth}{!}{\includegraphics[]{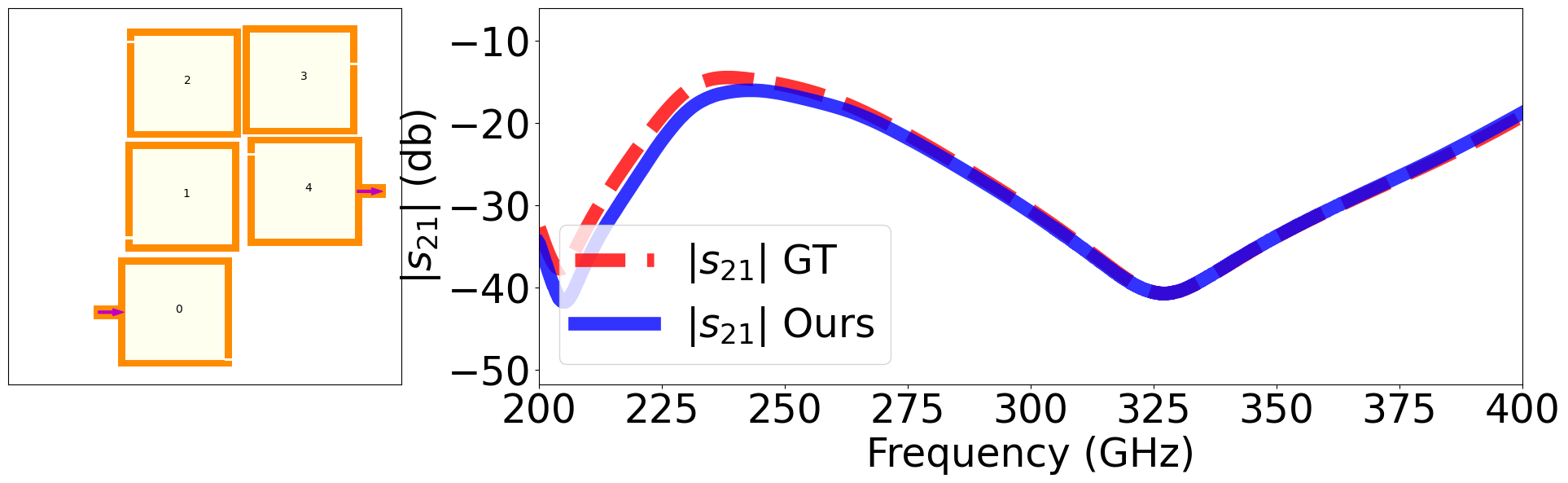}}
\end{minipage}%
\label{fig:12e}
}%
\subfigure[5-resonator circuit and its transfer function from \textbf{Circuit-GNN}]{
\begin{minipage}{0.5\linewidth}
\centering
\resizebox{\linewidth}{!}{\includegraphics[]{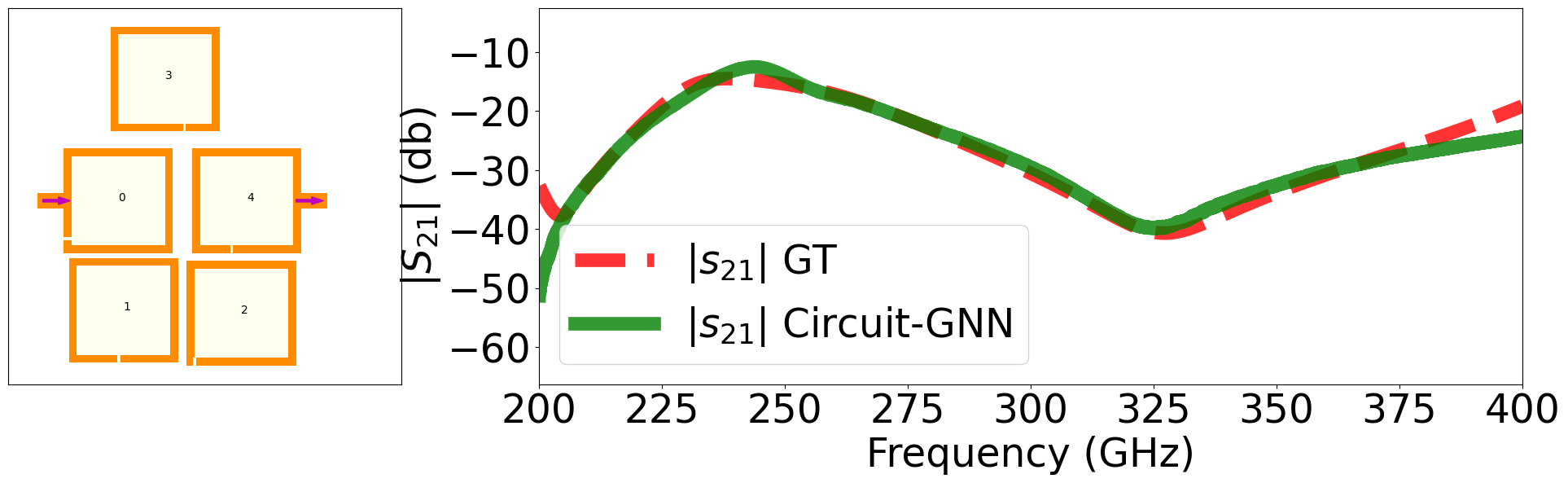}}
\end{minipage}%
\label{fig:12f}
}%

\subfigure[6-resonator circuit and its transfer function from \textbf{DCIDA}]{
\begin{minipage}{0.5\linewidth}
\centering
\resizebox{\linewidth}{!}{\includegraphics[]{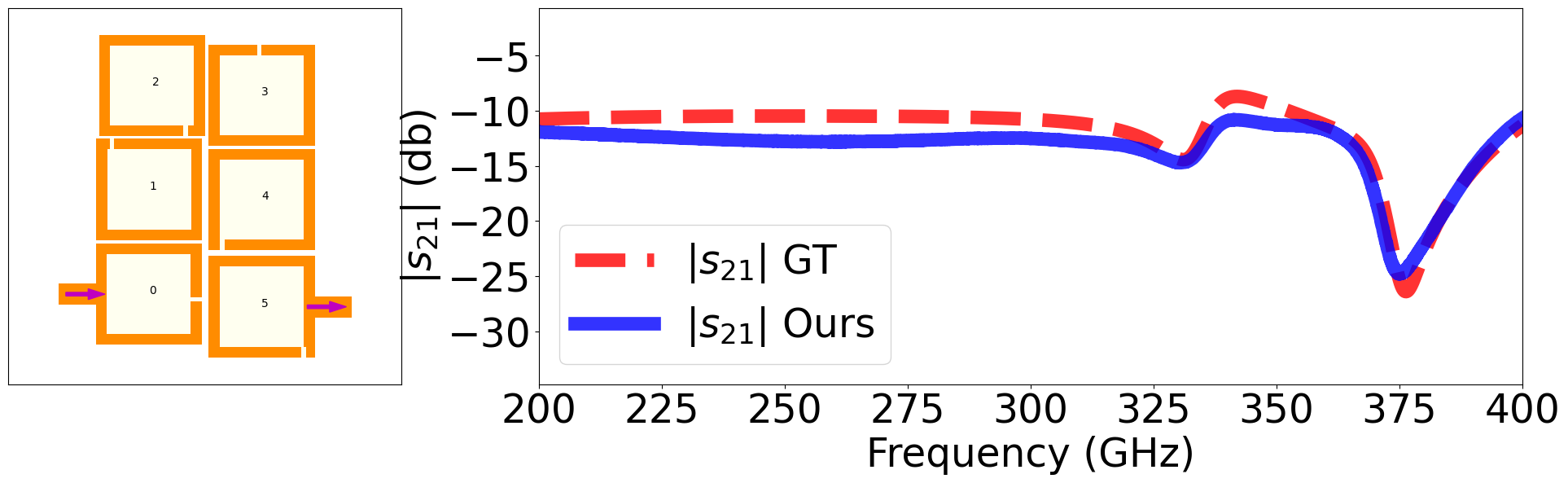}}
\end{minipage}%
\label{fig:12g}
}%
\subfigure[6-resonator circuit and its transfer function from \textbf{Circuit-GNN}]{
\begin{minipage}{0.5\linewidth}
\centering
\resizebox{\linewidth}{!}{\includegraphics[]{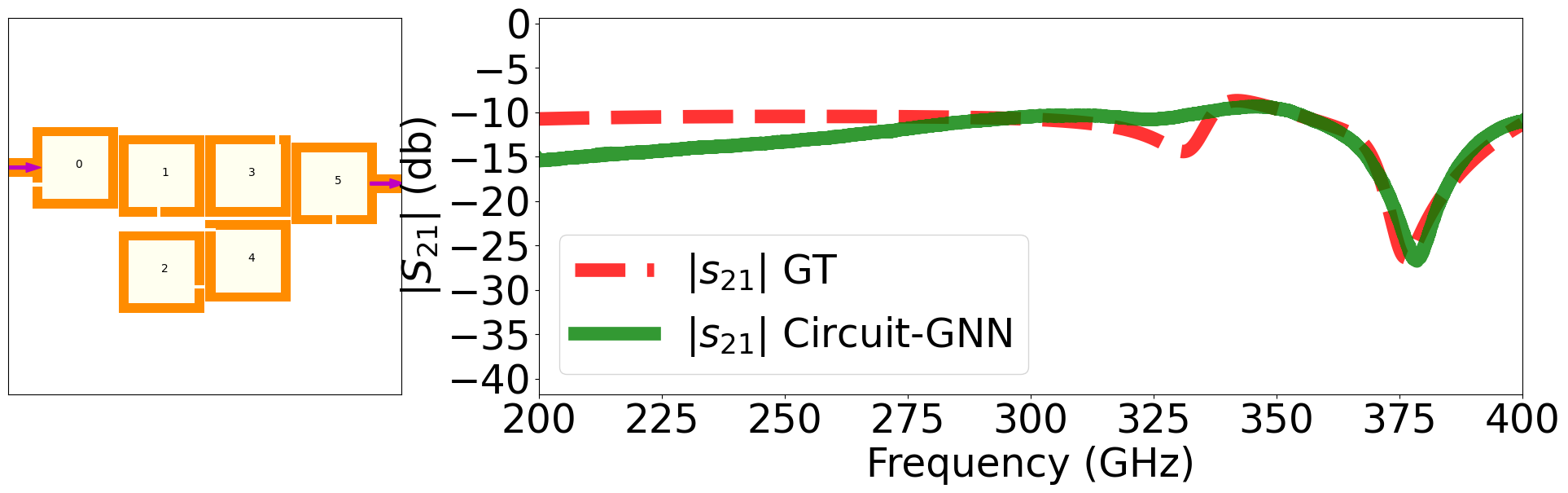}}
\end{minipage}%
\label{fig:12h}
}%
\centering
\vspace{-3mm}\caption{Visualization of inverse designs from Transformer-based DCIDA and Circuit-GNN with the \textcolor{red}{the challenging transfer functions as targets (ground truth (GT))}: we use blue color to show \textcolor{blue}{the transfer functions} of generated distributed circuits from Transformer-based DCIDA while applying green color to display \textcolor{Green}{the transfer functions} of generated distributed circuits from Circuit-GNN.} 
\label{fig12}
\end{figure*}

In the Figure \ref{fig:12a} and \ref{fig:12c}, the inverse designs from Transformer-based DCIDA have better transfer functions which tightly meet the target transfer function with two valleys and at least one peak, while the generated circuits from Circuit-GNN cannot have transfer functions fully meeting the trends of the target transfer functions. Even though the target transfer functions in Figure \ref{fig:12b} and \ref{fig:12d} has multiple kinks, inverse designs generated from Transformer-based DCIDA better satisfy the target transfer functions, when comparing with those produced by Circuit-GNN which cannot capture the trends of the kinks in the transfer functions.

\begin{figure*}[t]
\centering
\subfigure[3-resonator circuits]{
\begin{minipage}{0.25\linewidth}
\centering
\resizebox{\linewidth}{!}{\includegraphics[]{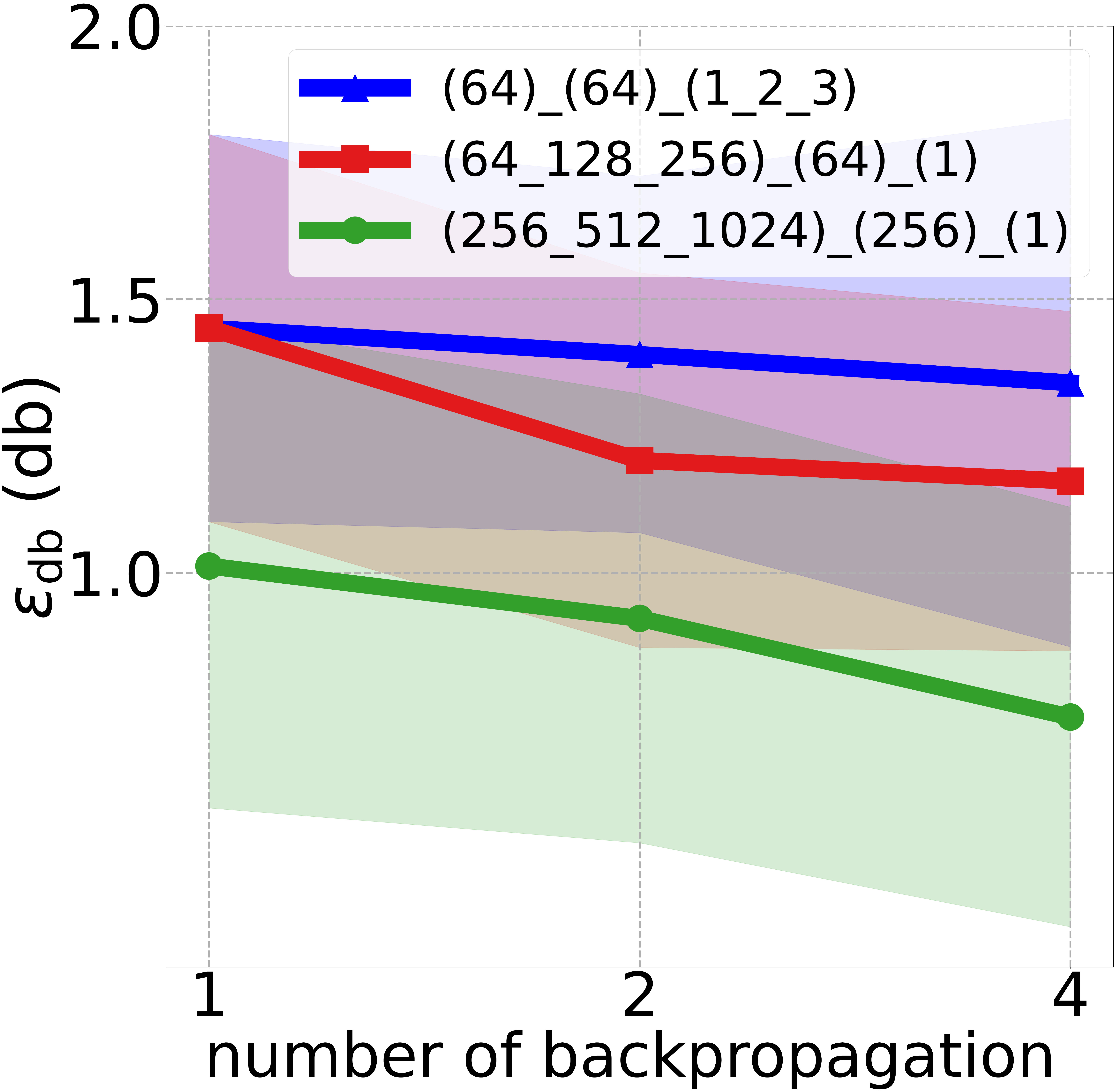}}
\end{minipage}%
\label{fig:13a}
}%
\subfigure[4-resonator circuits]{
\begin{minipage}{0.25\linewidth}
\centering
\resizebox{\linewidth}{!}{\includegraphics[]{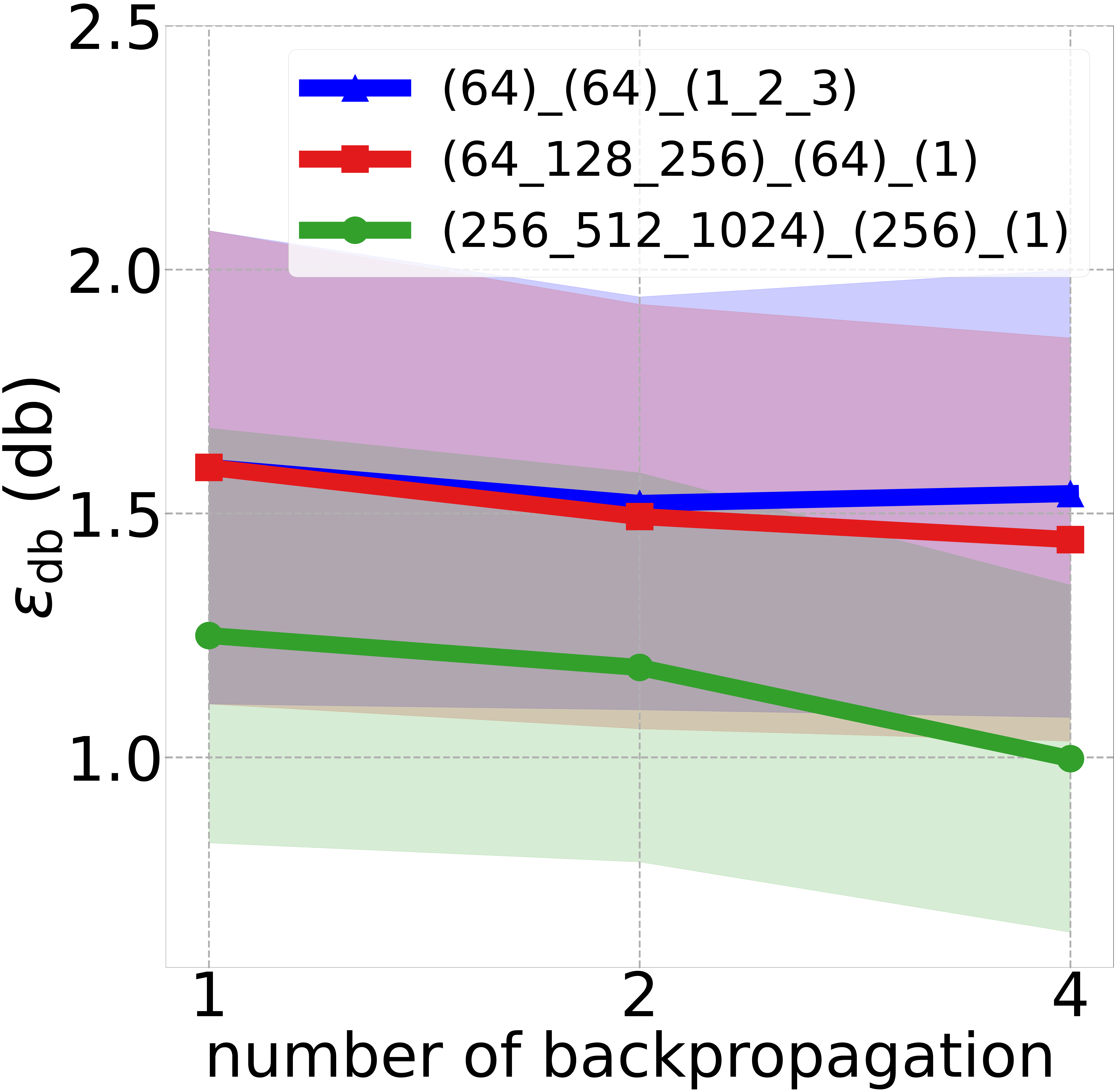}}
\end{minipage}%
\label{fig:13b}
}%
\subfigure[5-resonator circuits]{
\begin{minipage}{0.25\linewidth}
\centering
\resizebox{\linewidth}{!}{\includegraphics[]{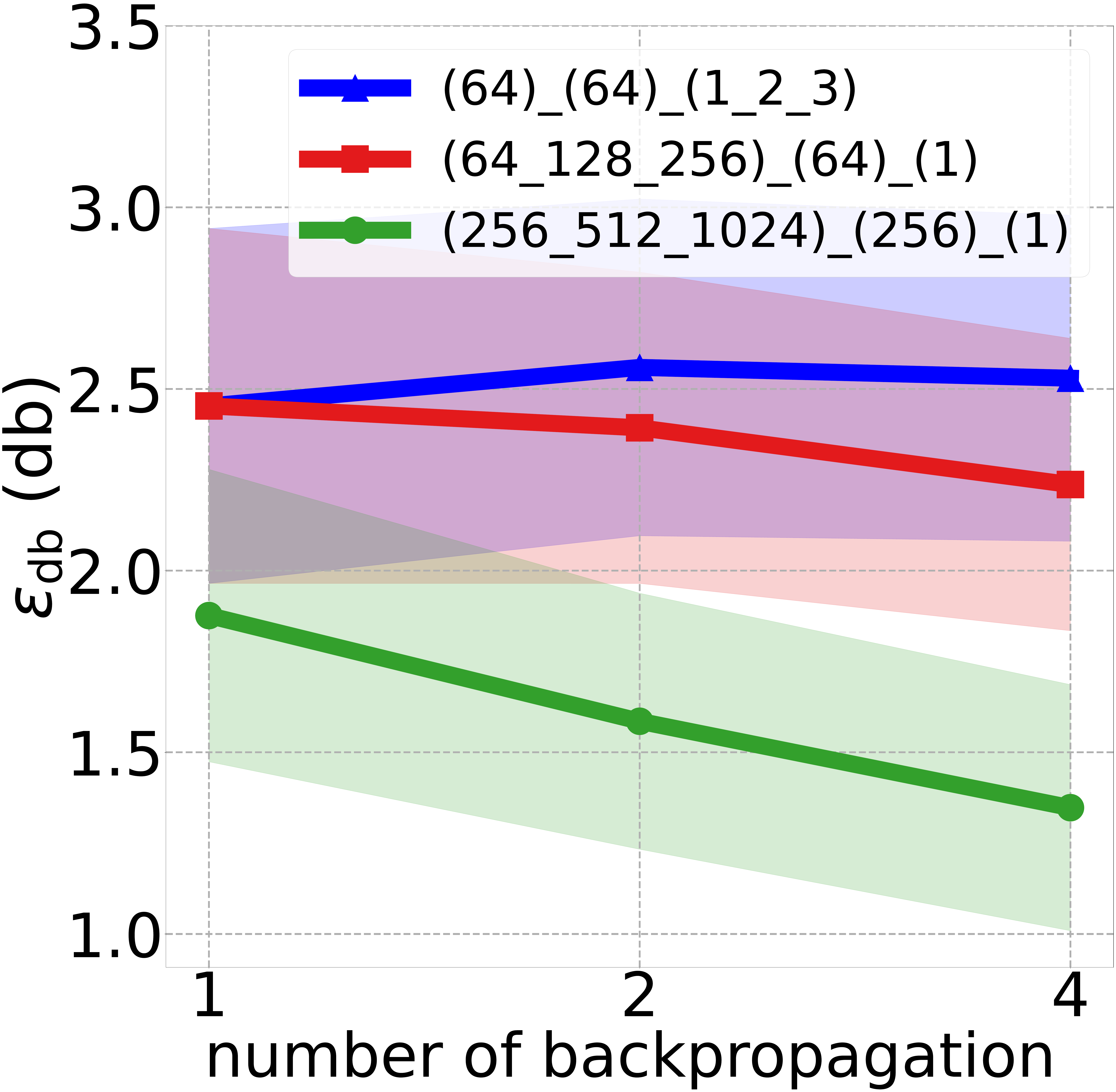}}
\end{minipage}%
\label{fig:13c}
}%
\subfigure[6-resonator circuits]{
\begin{minipage}{0.25\linewidth}
\centering
\resizebox{\linewidth}{!}{\includegraphics[]{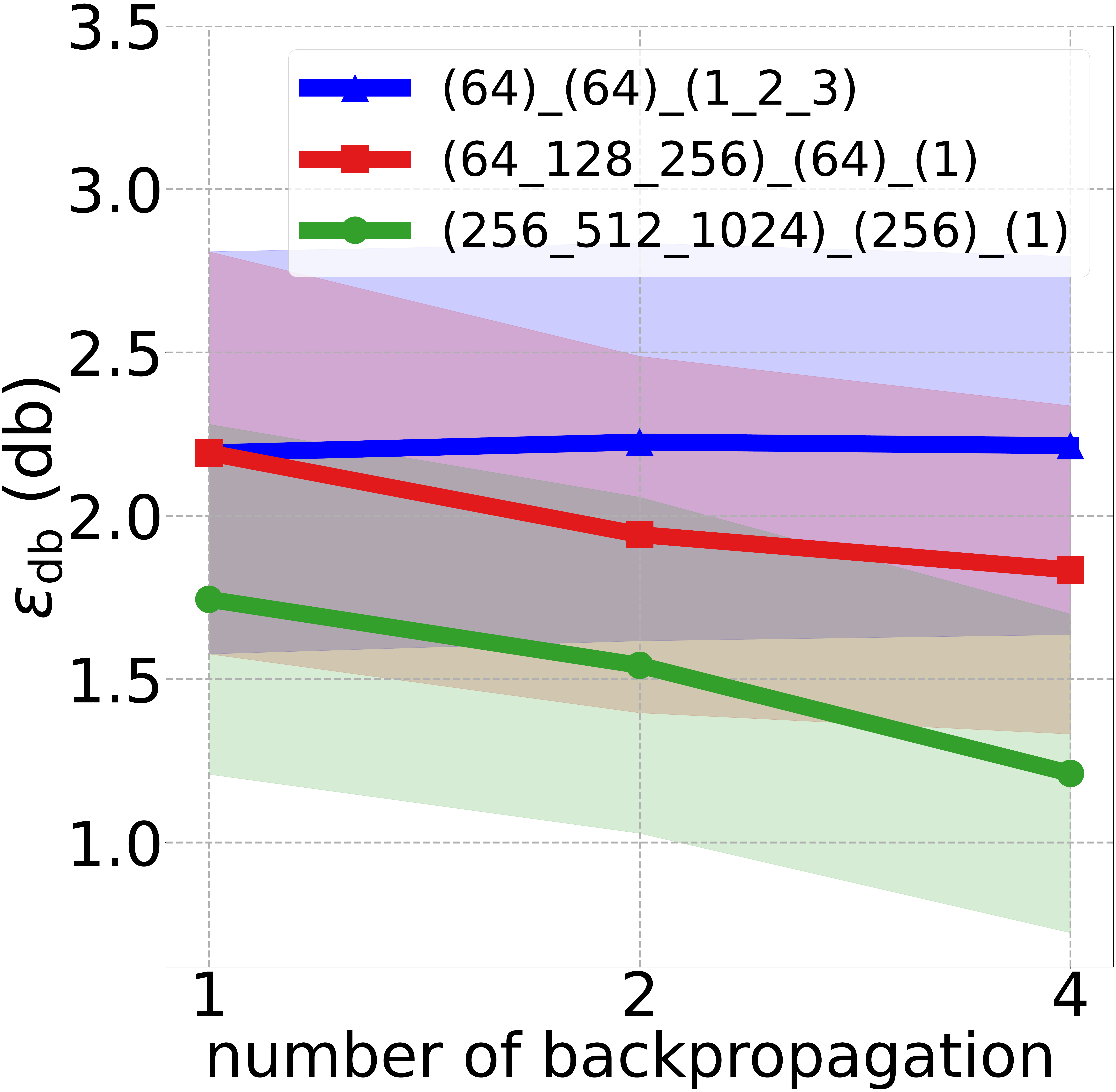}}
\end{minipage}%
\label{fig:13d}
}%
\centering
\vspace{-3mm}\caption{Error $\epsilon_\text{dB}$ of inverse designs from Transformer-based DCIDA varied with different loops of backpropagation, which are determined by batch sizes, mini batch sizes and number of epoch. We report average errors with standard deviations on circuits with different number of resonators} 
\label{fig13}
\end{figure*}

\section{Parameter Sensitivity Analysis}
\label{para_sensitivity}
\noindent \textbf{The effect of the batch size, the mini batch size and the epoch.} The number of loops of backpropagation are determined by $\frac{ZE}{z}$, where $Z$, $z$ and $E$ represent the batch size, the mini batch size and the number of epoch respectively. In the Figure \ref{fig13} based on different number of resonators circuits, we explore the effect of these variables on the performance of inverse designs from Transformer-based DCIDA. We define the format as ``(batch size)\_(mini batch size)\_(epoch)''. For example ``(256,512,1024)\_(256)\_(1)'' means three combinations of the variables including ``(256)\_(256)\_(1)'', ``(512)\_(256)\_(1)'' and ``(1024)\_(256)\_(1)''. The combinations corresponds to the number of loops of performing backpropagation once, twice and $4$ times due to $1=\frac{256*1}{256}$, $2=\frac{512*1}{256}$ and $4=\frac{1024*1}{256}$.

Figure \ref{fig13} shows that with the limited batch size and mini batch size (i.e., the batch size and the mini batch size are equal to 64.), even though we increase the number epoch leading to more loops of backpropagation, the improvement of performance of inverse designs from Transformer-based DCIDA are not significant. When we raise the batch size with keeping a smaller mini batch size and epoch equal to $1$, the error $\epsilon_\text{db}$ of inverse design from Transformer-based DCIDA slightly drops. As we set a larger batch size and a larger mini batch size while keeping the epoch equal to $1$, the performance of inverse design from Transformer-based DCIDA are significantly enhanced, accompanied by a sharp decrease in errors. We conclude that Transformer-based DCIDA with a larger batch size and a larger mini batch size can generate better inverse designs under the same number of loops of backpropagation.

\section{Algorithm}
\label{algo}
We summarize the proposed algorithm in Algorithm 1.
\begin{algorithm}[tb]
\caption{\underline{D}istributed \underline{C}ircuit \underline{I}nverse \underline{D}esign \underline{A}utomation (DCIDA) to generate distributed circuits}
\label{alg:clique}
\textbf{Input}: Target transfer function $\gY$, the number of resonators $N$, constant tensor $\mI$, number of iteration $t$, number of epoch $E$, batch size $Z$ and mini batch size $z$, a simulator or an approximator of a simulator $\hat{\gY}$;\\
\textbf{Output}: Generative circuit $\xi$.
\begin{algorithmic}[1] %
\STATE Initialize a neural network $Net_{\theta}$ with initial parameters $\theta$ and $N$
\FOR{$i = 1$ to $t$}
\STATE $\gF \leftarrow Net_{\theta}(\mI)$
\STATE $\gA \leftarrow Sample(\gF)$ with $Z$ times
\STATE $\xi_b \leftarrow Map(\gA)$ with $Z$ batches
\STATE $\mR \leftarrow Compute\_Reward(\xi_b)$ with $\gY$ and $\hat{\gY}$ based on Eq.(\ref{db}) and Eq.(\ref{eq3})
\STATE Save the best inverse design $\xi$ with the best $R\in \mR$.
\STATE $\hat{R} \leftarrow Compute\_Running\_Reward(\mR)$
\FOR{$j=1$ to $\frac{ZE}{z}$}
\STATE $\gF^{\prime} \leftarrow Net_{\theta}(\mI)$
\STATE $A \leftarrow Compute\_Advantage(\hat{R}, \mR)$
\STATE $\gL(\theta) \leftarrow Compute\_Loss(\gF, \gF^{\prime}, \gA, A)$ based on Eq.(\ref{loss})
\STATE $\theta \leftarrow Optimize(\gL(\theta))$
\ENDFOR
\ENDFOR
\STATE Obtain the generative circuit $\xi$
\end{algorithmic}
\end{algorithm}\vspace{-2mm}

\end{document}